\documentclass[10pt,letterpaper]{article}

\usepackage[utf8]{inputenc}
\usepackage[margin=0.75in]{geometry} 
\usepackage{amsmath,amssymb,amsfonts} 
\usepackage{graphicx}                
\usepackage{booktabs}                
\usepackage{hyperref}                
\usepackage{microtype}               
\usepackage{cite}                    

\newcommand{\thetastart}{\theta^-}
\newcommand{\thetastop}{\theta^+}
\newcommand{\thetav}{\tilde{\theta}}

\newcommand{\qv}{\tilde{q}}
\newcommand{\weight}{\omega}
\newcommand{\bigoh}{\mathcal{O}}
\newcommand{\summ}{\mathfrak{m}}
\newcommand{\sumn}{\mathfrak{j}}
\newcommand{\suml}{\mathfrak{j}}

\newcommand{\Rbar}{{R^*}}
\newcommand{\thetabar}{{\Theta^*}}
\newcommand{\Krate}{\kappa}
\newcommand{\thetaobs}{\phi}
\newcommand{\coupling}{\mathcal{N}}
\newcommand{\Real}{\,\textbf{Re}}
\newcommand{\bearing}{\Phi}
\DeclareMathOperator{\Arg}{Arg}

\newcommand{\ajb}[1]{\textcolor{red} {AJB: {#1}}}

\hypersetup{
    colorlinks=true,
    linkcolor=blue,
    citecolor=blue,
    urlcolor=blue
}

\title{A weighted model of perception and decision-making \\ 
between targets of finite size}

\author{
  \textbf{W. Christopher Strickland} \\
Department of Mathematics,\\
Department of Ecology and Evolutionary Biology,\\
National Institute for Modeling Biological Systems (NIMBioS), \\
University of Tennessee Knoxville, Knoxville, TN, USA \\
  \texttt{cstric12@utk.edu}
  \and
  \textbf{Andrew J. Bernoff} \\
 Department of Mathematics,\\
 Harvey Mudd College, Claremont, CA, USA \\
  \texttt{bernoff@g.hmc.edu}
}

\date{\vspace{-5ex}} 

\begin{document}

\maketitle

\begin{abstract}
Recent research grounded in experiments has connected neural ring models of vision to how animals navigate a complex landscape of attractive targets. In this paper we investigate the mathematical and biological implications of a three-stage model where animals pre-process visual stimuli to identify a discrete set of targets, process this input to select the dominant targets, and then post-process this information to navigate the landscape. Incorporating finite target sizes and a neural density allows consistent target acquisition and pursuit reflecting what is seen in nature. Mathematically, we show this model corresponds to an energy minimization problem, simplifying both its analysis and numerical implementation. Biologically, we argue that the model is physiologically motivated, reproduces experimental observations, and presents a fairly direct pathway to decoding neural geometry via empirical data. Our results also demonstrate an analysis pipeline for future model extensions and suggest specific avenues for experimental research to test and parameterize our quantitative framework.
\end{abstract}

\bigskip

 \textbf{Subject:} Mathematical biology, agent-based modeling, collective behavior, movement ecology, navigation

\bigskip


\newpage
\section{Introduction}

Agent-based models (ABMs) of collective motion and spatial decision-making have historically relied on simplified assumptions regarding individual perceptual capabilities. Standard approaches frequently model agents with omniscient information, assuming individuals possess global access to the locations and states of all nearby neighbors or targets regardless of occlusion, spatial extent, or visual processing constraints \cite{Sum2005,CouKra2003,VicZaf2012}. 
While more recent formulations, such as the work of Bastien \& Romanczuk \cite{BasRom2020} and Kronguez, Ayali, \& Kaminka \cite{KroAyaKam2024} have begun addressing how limited field-of-view and visual occlusion may govern local interactions, these models essentially try to reverse engineer field observations to obtain a plausible parameterized set of models and then examine the dynamical collective behavior observed. 

Recent work by Sayin et al \cite{SayCouPet2025} uses field, laboratory, and virtual reality experiments to argue that locusts do not explicitly align with their neighbors and suggest that they identify the bearings to conspecifics and then reach an internal consensus from this information for a direction of motion. This perspective supports a modeling framework concentrating on how an individual agent reacts to visual stimuli and builds on a set of models developed by Dr. Iain Couzin and collaborators grounded in an ego-centric view of decision-making \cite{Sridhar2021,Oscar2023,Gorbonos2024}. Models of this type decompose sensory-driven navigation into three distinct phases:
\begin{enumerate}
    \item \textbf{Pre-processing:} Visual and spatial inputs are reduced to a discrete or continuous set of relevant targets.
    \item \textbf{Consensus processing:} Neural dynamics integrate competing sensory signals to produce a single neural consensus vector for a direction of motion.
    \item \textbf{Post-processing:} The observer translates the neural consensus vector into physical steering and motor control to navigate through space.
\end{enumerate}
To capture the consensus phase, neural ring models and spin systems---most notably the Ising model---have proven effective. Incorporating neural and environmental noise into this architectures captures how noise informs state transitions and decision-making. Crucially, lateral inhibition within these networks leads to the suppression of peripheral signals, allowing animals to choose between competing visual targets rather than averaging them into compromised trajectories. 

In this paper, we extend and refine this perception-to-navigation pipeline to develop a computationally efficient, biologically faithful model of spatial decision-making between finite-sized targets. We relax the point-source assumption by introducing a pre-processing framework that explicitly accounts for targets with non-zero spatial footprint and finite visual angles.  It is well established that visual acuity is best near the center of the visual field and diminishes significantly in the periphery \cite{KupBenCar2022,Muller2005,Reynolds2009}. To model this effect we introduce the concept of a neural mapping that warps the visual field and favors targets in the fore over the periphery. We then model consensus processing by incorporating neural noise via Glauber dynamics \cite{Gla1963} into the Ising spin framework (a mild extension of the work of \cite{Gorbonos2024}). This framework allows us to characterize the consensus process as a minimization problem in an energy landscape where local minimizers represent potential bearings and transition of state stability govern target selection.    
    
Mathematically, our formulation of neural dynamics allows us to embed the discrete spin system into a continuous framework as a minimization problem for a quadratic functional which admits a spectral decomposition. Naively the Ising model of neural dynamics requires an all-to-all coupling of $\mathcal{O}(N^2)$ complexity (where $N$ is the number of active neurons); we show that our warping allows a rank two spectral reduction which has a complexity of $\mathcal{O}(2N)$ and seems more biologically plausible. Our neural mapping also reproduces the classical \emph{mexican hat} model of lateral inhibition \cite{Muller2005}, where a target in the fore of the visual field suppresses peripheral visual signals to ensure decisive target selection. Finally, we introduce a new, kinematically consistent steering law where the walker's speed is constant but the angular turning response increases monotonically as the angle between the walker's bearing and the neural consensus vector increases.

Applying the two pre-processing concepts of neural warping and neural weighting, we analyze the resulting bifurcation geometry for spatial decision-making and its implication for noisy walkers attracted to either $\delta$-function targets or targets of finite extent. We first demonstrate that $\delta$-function targets are a singular case: they do not represent limiting behavior as target size shrinks to a point. We then show that for observers without a rear-facing blind-spot, neural weighting based on a target's visual extent produces a defined region where an observer is always attracted to the near target regardless of initial heading. If neural weighting is additionally made to bias the observer toward front-appearing targets, this region disappears.

Our results illustrate that our modeling framework for finite-sized targets can reproduce two-target bifurcation structures seen in published experimental data. However, with three targets, the bifurcation structure changes versus what has been previously demonstrated for $\delta$-functions. Instead of a bifurcation cascade that cleanly subdivides decision-making into smaller groups, we show that the bifurcation structure instead collapses back onto the center target, leaving the outer targets to become stable only in separate, detached saddle-node bifurcations. Consequently, while noise can recover the expected 1:2:1 decision ratio between targets of finite extent, a walker moving with only a small amount of angular noise will be heavily biased toward the middle target.

This work establishes a strong mathematical basis for further exploration into the spatial decision-making patterns we see both experimentally and in nature. In particular, our work demonstrates how robust the neural ring model framework is, even within a low-dimensional setting, and it opens the door for both further experimental work to test predictions and further mathematical work to refine the model while expanding the varieties of external stimuli we can consider.
    
\section{Methods}

  \subsection{Perception-based neural model}

Consider a brain characterized by a 
neural network composed of $N$ spins (neurons) that receive stimuli from a collection of $K$ targets; spins not receiving stimuli from a target are neglected. We group the spins by the target stimulating them, thereby resulting in $K$ non-overlapping groups of spins which we will denote $\{G_1, G_2, \cdots G_K \}$. We will assume that each spin exists in one of two states: $\sigma_j=0$ or $\sigma_j=1$ - e.g., a neuron that is not firing or firing.

In a departure from \cite{Sridhar2021}, the number of spins in each group will generally be different, and we assume that their relative sizes are a function of three different key factors. First, the angular extent of the target affects group size, with larger targets stimulating more neurons. Second, the neural band may be nonuniform in such a way that there are more neurons associated with some areas of the visual field than others. This implies that a target in front of an observer could be associated with a bigger neural group than one that is the same distance behind or off to the side. Third, each target has an associated signal strength, $s_k$, which directly affects group size. This allows less attractive targets to generate a lower signal strength and therefore a smaller group size which enables us to incorporate preferential decision-making among potential target choices including, for example, pursuing a food source versus a conspecific.

The neural model has two stages: a pre-processing stage that determines the size of the neural group stimulated by each target and a processing stage where the spins interact and the resulting dynamics select which neurons are excited.  Eventually the excited neurons determine a \emph{neural bearing} which dictates navigational decisions.

\subsection{Target perception and neuron group size}
For simplicity, we confine ourselves to a two-dimensional plane and consider an egocentric (observer-based) coordinate system with our observer at the origin facing along the polar axis. The observer perceives targets at egocentric angles, $\theta\in[-\pi,\pi]$, with no depth perception and (at this point) no blind spot; we will assume $\theta=0$ corresponds to the observer's direction of motion.

Scattered around the observer are $K$ targets (enumerated by $k \in [1,2,\cdots,K])$ which are assumed to be bounded regions or $\delta$-functions. For simplicity, we will assume that all targets are circular with fixed, identical radius. Since the observer has no depth perception, targets are perceived as occupying a sector.
To each target, we also assign a signal strength, $s_k \in [0,1]$ which reflects its 
intrinsic attractiveness to the observer. The observer then sees the angular extent of each target based upon its apparent width from the observer's position in space. This angular width is a function of the target's physical size, its distance to the observer, and whether or not the target is partially obscured by other objects in front of it. 

In this study we concentrate on configurations where targets are circles of equal size and whose signal strength is unity ($s_k=1$). This simplifies the occlusion problem as there cannot be smaller targets obscuring only an interior portion of a larger target that is further away. As such, target $k$ generally appears to the observer as an interval $[\thetastart_{k},\thetastop_{k}]$; for $\delta$-functions $\thetastart_k=\thetastop_k$. The one exception is the case where a target is directly behind the observer and therefore occupies $[-\pi,\thetastart] \cup [\thetastop,\pi]$, but this case can be treated analogously. 

Important to the biological interpretation of our results is the fact that for non-obscured circular targets in 2D, distance can be determined from only the target's radius and apparent visual extent as seen from the observer. In 3D space with targets that have more complicated geometries, vertical visual extent may be used for this purpose instead, with experimental work suggesting this may play an important role in locust behavior \cite{BleSheKam2024}. It is straightforward to extend our framework to targets of any connected geometry and variable size by considering visible sets associated with each target rather than closed intervals. Details on how to reduce a two-dimensional array of targets, some of which may be occluded or obscured, to a collection of sectors has been addressed before \cite{BasRom2020,KroAyaKam2024}.

\subsubsection{Target perception}

Following previous work, we will associate to the $k^{th}$ visible target an angular location ($\theta_k$) and a group of neurons ($G_k$ with cardinality $|G_k|$) stimulated by the target. First, we take $\theta_k$ to be the egocentric angle to the midpoint of the $k^{th}$ target, regardless of if the midpoint is actually visible. This decision is based on the assumption that the observer can infer the center point of targets from limited information \cite{KroAyaKam2024}. 
The cardinality of neural group $G_k$ is then computed from the visible extent of the target $[\thetastart_k,\thetastop_k]$ using the neural density. 

Consider a neural weighting $\weight(\theta)$ and define the number of neurons in the $k^{th}$ group as
\begin{equation}
|G_k| = s_k\int_{\thetastart_k}^{\thetastop_k}
\weight(\theta)\ d\theta\label{eq:group_size}
\end{equation}
where $s_k$ is a signal strength reflecting the target's attractiveness (here taken to be 1).  We consider two choices for the neural weighting (i) \textbf{weighted} where $\weight(\theta)$ is equal to the neural density $\mu(\theta)$ or another function, or (ii) \textbf{unweighted} where $\weight(\theta) = 1/2\pi$. 
Either formulation weighs targets with larger visual extent (e.g., nearby targets) higher than those that appear smaller. The first model can bias the observer toward targets that are toward the front of its visual field (a foveal effect) where the neural density is highest. 

In the case of $\delta$-function targets, we instead take $|G_k| = s_k \weight(\theta_k)$. Note that since $|G_k|$ is the number of neurons associated with the $k^{th}$ target it should be an integer. However, our interest lies in the relative size of different groups when the total number of spinners becomes large, and so this distinction will shortly become negligible.

\subsubsection{The neural mapping}

As we know from our own vision, perception is neither uniform nor omnidirectional \cite{KupBenCar2022}. Moreover, recent studies suggest that anisotropic perception favoring forward vision is necessary for aligned swarms \cite{GaoCou2025}.
A key feature of our approach is that we will model this non-uniformity via a \emph{warping} of angular perception which we will refer to as a \emph{neural map}. 

Assume that the observer can see targets in some symmetric interval $\theta \in [-b,b]$ where $b$ is taken as a positive constant with $b<\pi$ reflecting a blind spot behind the observer. The neural map $\thetav =\mathcal{U}(\theta)$ is an  anti-symmetric function that maps individual angles $\theta \in [-b,b]$ to $\thetav \in [-\pi,\pi]$. 
We construct this mapping by deriving it from a \textit{neural density function} $\mu(\theta)$ which reflects variations in the distribution of neurons which typically are concentrated in the fore and decrease as one approaches the periphery. We choose this density to be symmetric, $\mu(-\theta)=\mu(\theta)$, and to have unit mass,

$$\int_{-\pi}^{\pi} \mu(\theta) \, d \theta = 1, $$
and then define the neural map as a scalar multiple of the number of neurons between the fore and some angle $\theta$ in the visual field,
\begin{equation}
\thetav =
\mathcal{U}(\theta)  \equiv { 2 \pi}\int_0^{\theta}\mu(\phi)\ d\phi.\label{eq:angle_map}
\end{equation}
where the mapped \textit{neural angles} $\thetav$ are in $[- \pi, \pi]$.

In this setting, a larger neural density shifts the resulting neural angles further away from the front of the visual field, spreading those angles apart, while lower neural density will result in more compacted angles. In general, we will assume that $\mu$ is continuous and decreases to zero as $|\theta|$ increases from $0$ to $b$.  A classic example is a piece-wise linear function
\begin{equation}
\label{eq:linearwarp}
\mu(x) = 
\begin{cases}
	  \frac{1}{a+b} & 0 \le |x| \le a, \\
	\frac{1}{a+b} \left [\frac{b-|x|}{b-a} \right ]  & a \le |x| \le b , \\
    0 & b \le |x| \le \pi
\end{cases}
\end{equation}
for $0 \le a < b \le \pi $.
In practice, one could choose any anti-symmetric, one-to-one function, $  \ \mathcal{U}:[-b,b]\rightarrow[- \pi,\pi]$ that maps the visual range onto $[-\pi,\pi]$ for the mapping of individual angular positions to neural positions. Our choice here is for simplicity; more realistic models could be chosen based on studies in the biological literature (see for example \cite{KraGab2005,KupBenCar2022}).

We contrast this with previous work \cite{Sridhar2021, Oscar2023, Gorbonos2024} which look at angular distortions where the perceived \emph{distances between targets} is a function of the angular difference between their locations, that is  $$d(\theta_1,\theta_2) = \mathcal{I}(\theta_1-\theta_2) $$
for some function $\mathcal{I}$. This approach is discussed and contrasted with the warpings described above in Section \ref{sec:coupling_kernel}.


\subsection{Neural dynamics}
\label{subsec:ND}

The goal now is to take our model of visual perception and neural activity and produce a decision about where the observer would like to go. We will formulate this process as an energy minimization procedure that occurs among neurons within the observer's brain. Our basic assumption for this process, following the work of \cite{Sridhar2021}, is that contributions to the energy follow an Ising model with every pair of neurons coupled within an energy that is pairwise and additive. Naively, this assumption is unbiological - if there are $N$ neurons there must be $\bigoh(N^2)$ couplings which seems unlikely when $N\approx10^5-10^6$ for a typical insect \cite{matsliah2024neuronal,KraGab2005}. However, we will demonstrate that this number can be reduced to $\bigoh(2N)$ for certain types of neural couplings.

Let $N=\sum_{k=1}^K |G_k|$ be the number of stimulated neurons. The $\summ^{th}$ neuron has associated to it a binary spinor, $\sigma_\summ \in \{0,1\}$, where a value of 0 indicates the spinor is \emph{off} (neuron is not firing) and 1 indicates the spinor is \emph{on} (neuron is firing). Neural biology has established that when a neuron fires it can either enhance or suppress the likelihood that another neuron will be excited depending on the neural geometry. Our assumption is that the pairwise energy depends solely on the neural angles $(\thetav_\summ, \thetav_\sumn)$ associated with the pair of spinors $(\sigma_\summ,\sigma_\sumn)$ and that this energy is additive. 

The Hamiltonian is then given by
\begin{equation}
\mathcal{H} = -\frac{\mathcal{E}}{N}\sum_{\summ=1}^N \sum_{\substack{\sumn=1,\\ \sumn \ne \summ}}^N\sigma_\summ \sigma_\sumn J(\thetav_\summ,\thetav_\sumn),\label{eq:H_orig}
\end{equation}
with interaction kernel $J(\thetav_\summ,\thetav_\summ)$ defining the pairwise energy contribution of neurons at $\thetav_\summ$ and $\thetav_\sumn$. Pairs of \emph{on} spins with positive $J(\thetav_\summ,\thetav_\sumn)$ lower the overall energy (an excitatory interaction) while positive $J(\thetav_\summ,\thetav_\sumn)$ values raise the energy (an inhibitory interaction). We normalize $J$ so that $J(\thetav,\thetav)=1$. The constant $\mathcal{E}$ has units of energy. Previous work \cite{Sridhar2021,Gorbonos2024} scaled the Hamiltonian with the number of targets, an assumption that follows logically only when targets are identical. Eventually we will consider this sum in the limit of large $N$ and the prefactor of $1/N$ ensures that the energy scales proportional to the number of neurons.

We now reformulate the energy by grouping spinors associated with each target. Let $\rho_k=|G_k|/N$ be the fraction of the total number of spinors associated with the $k^{th}$ target and define the state vector $\vec{n}$ as
\begin{equation}
\vec{n} = \langle n_1, n_2, \cdots n_K \rangle  \ , \qquad 
n_k = \frac{1}{N}\sum_{\sigma_\ell\in G_k} \sigma_\ell \ .
\label{eq:nvec}
\end{equation} 
Here $n_k$  is the fraction of the total number of spinors that are both \emph{on} (firing)  and in group $G_k$. 
Adding back and subtracting the terms where $\summ=\sumn$ in Eqn. \ref{eq:H_orig}, we can reformulate the Hamiltonian by summing over groups of spinors associated with the same target
\begin{align}
    \mathcal{H} &= -\frac{\mathcal{E}}{N} \left [ \sum_{k=1}^K\sum_{\ell=1}^K (Nn_k) (Nn_\ell) J(\thetav_k,\thetav_\ell) -\sum_{k=1}^K (Nn_k) \right ]\nonumber\\
    &= -{\mathcal{E} N} \sum_{k=1}^K n_k 
    \left(\sum_{\ell=1}^K n_\ell J(\thetav_k,\thetav_\ell)-\frac{1}{N}\right).\label{eq:H_rewrite}
\end{align}
We now pass to a continuum limit by letting $N$ tend to infinity. In this limit we see that 
\begin{equation}
\mathcal{H} = \mathcal{E} N\left [ 
H(\vec{n}) + \bigoh \left ( \frac 1 N \right ) 
\right] , \qquad 
H(\vec{n}) = - \sum_{k=1}^K \sum_{\ell=1}^K n_k n_\ell J(\thetav_k,\thetav_\ell) \ ,
\label{eq:H}
\end{equation}
where $n_k$ are now continuous variables in a rectangular parallelepiped $\mathcal{P}$ defined by $ 0 \le n_k \le \rho_k=|G_k|/N$. 
The $\bigoh(1/N)$ error term reflects the sub-dominant self-interactions in $\mathcal{H}$.
\footnote{The reader is cautioned that for notational convenience we may refer to the energy $H$ as $H(\vec{n})$ or $H(n_k)$ when we wish to emphasize the dependence on all or one of the state variables.}

\subsection{Neural noise, thermal fluctuations, and Glauber dynamics}
\label{subsec:Glauber}
In the absence of neural noise, one can minimize $H$ and the normalization $J(\thetav,\thetav)=1$ guarantees that the critical point at $\vec{n}=\vec{0}$ is not a minimizer; therefore, the mimimizer will occur somewhere on the boundary of $\mathcal{P}$ where the KKT conditions are satisfied \cite{BoyVan2004}.
However, neural systems are inherently noisy. Following the work of many others we model this as a Markov process \cite{Gla1963}: neurons will transition randomly between excited and quiescent states at a rate that depends upon the energy change induced by the transition. We assume that these transitions follow Boltzmann statistics \cite{Gla1963,New1999}.  

First, we calculate the energy change associated with a single spinor in the $k^{th}$ group being activated. Eqn. \eqref{eq:nvec} shows that this corresponds to $n_k \to n_k + \tfrac 1 N$, so the change in the dimensional energy is
$$ (\Delta \mathcal{H})_k = 
\mathcal{E} N H \left ( n_k + \tfrac 1 N\right ) -\mathcal{E} N H \left ( n_k \right ) \approx {\mathcal{E}} 
\frac{\partial H}{\partial n_k} \ .
$$
Glauber \cite{Gla1963,New1999} proposed that the transition rates for a spinor whose quiescent and active states differ by $\Delta \mathcal{H}$ are 
\[
r_{0\rightarrow 1} = \frac{1}{\tau_0} \frac 1 {1 + e^{ \Delta \mathcal{H} / k_B T}} \qquad \text{and}
\qquad 
r_{1\rightarrow 0} = \frac{1}{\tau_0} \frac 1 {1 + e^{ -\Delta \mathcal{H} / k_B T}}
\]
where $\tau_0$ is a timescale associated with the neural noise, $k_B$ is the Boltzmann constant, and $T$ is the absolute temperature (in Kelvin). The population of excited neurons in group $G_j$ is proportional to $n_j$ and the number of quiescent neurons is proportional to 
$\rho_j-n_j$, so the evolution equation for $n_j$ under Glauber dynamics can be written as
\begin{align}
\frac{dn_j}{dt} &= \frac{\rho_j-n_j}{\tau_0 
\left [1+\exp\left((\Delta \mathcal{H})_j / k_B T\right) \right ]} - \frac{n_j}{\tau_0 
\left [ 1+\exp\left(-(\Delta \mathcal{H})_j / k_B T\right) \right ]}
\\
&=
\frac{1}{\tau_0}
\left [ 
\frac{\rho_j}
{1+\exp{\left((\Delta \mathcal{H})_j / k_B T \right)} }
- n_j 
\right ].
\end{align}
which is equivalent up to a scaling to $dn_j/dt$ in Sridhar et al. \cite{Sridhar2021} and Gorbonos et al. \cite{Gorbonos2024}.
Now, if we introduce a timescale $\tau =t/\tau_0$ and define a noise parameter
$$\beta = \frac{\mathcal{E}}{k_BT} $$
the equation reduces to a set of $K$ ODEs, one for each target,
\begin{equation}
\frac{dn_k}{d\tau} = F_k(\vec{n}) \ , \quad F_k(\vec{n}) = \frac{\rho_k}
{1+\exp{\left( \beta (\Delta E)_k \right)} }
- n_k  \ , \quad (\Delta E)_k = \frac{\partial H}{\partial n_k}  \qquad \textrm{for} \ k=1,2, \cdots, K \ .    
\label{eq:glauberODE}
\end{equation}
Equilibrium solutions, $\vec{n} = \vec{n}^*$, satisfy the implicit equations 
\begin{equation}
	n_k^* = \frac{\rho_k}
{1+\exp{\left( \beta (\Delta E)_k^* \right)} } \ , \qquad (\Delta E)_k^* =
\left . \frac{\partial H}{\partial n_k} \right |_{\vec{n}=\vec{n}^*}
= - 2  \sum_{l=1}^K  n_\ell^* J(\thetav_k,\thetav_\ell).
\label{eq:steadystate}
\end{equation}
These equilibrium solutions must be in the interior of $\mathcal{P}$ (as $\exp(\beta (\Delta E)_k^*) >0$). In Appendix \ref{app:ndstability} we show that this system has a Lyapunov functional which guarantees the existence of at least one stable equilibrium although often multiple stable equilibria exist.

\subsection{The Coupling Kernel: Striving for Low Dimensionality}
\label{sec:coupling_kernel}

A drawback of the Ising model is that it is \emph{all-to-all} coupling: if there are $N$ neurons, we postulate $\mathcal{O} (N^2)$ neuronal couplings, which scales unbiologically. By contrast, for the commonly used cosine coupling \cite{Sridhar2021}, 
\begin{equation}
\label{eq:cosineneural}
J(\phi,\phi') = \cos(\phi-\phi') \ ,
\end{equation}
we show that the problem can be reduced to $\mathcal{O} (2N)$ couplings. We also describe a strategy for engineering low-rank kernels with desirable characteristics and describe a method for approximating a general (full-rank) kernel with a lower rank kernel of dimension $D$.

Appendix \ref{app:coupling} contains a description of how to identify the rank, $D$, of a general coupling kernel $J(\phi,\phi')$ and how that leads to $\mathcal{O} (D\cdot N)$ couplings. The rank of the cosine coupling is two, which is arguably the minimal rank coupling one can use for navigation (see Appendix \ref{app:cosine_kernel}). Furthermore, the neural mapping $\thetav=\mathcal{U}(\theta)$ generates a family of rank two kernels from the cosine kernel. In essence, since
$$J(\theta , \theta')=\cos(\thetav - \thetav') = \cos(\mathcal{U}(\theta)-\mathcal{U}(\theta'))$$
the spectral decomposition of the kernel in the warped coordinates yields the analogous decomposition in the unwarped coordinates,
\begin{equation}
J(\theta , \theta') = Q^+(\theta)Q^-(\theta') +Q^-(\theta)Q^-(\theta') \ . 
\label{eq:ranktwo}
\end{equation}
with $Q^+(\theta)=\cos(\mathcal{U}(\theta))$ and $Q^-(\theta) = \sin (\mathcal{U}(\theta))$. The takeaway message here is that the rank of the kernel is preserved under the warping of the coordinate system.

As the warping function here is arbitrary (up to some constraints) this generates an infinite family of rank two kernels.  One can generate an even larger family by choosing an even [$Q^+(\theta)$] and an odd [$Q^-(\theta)$] pair of functions (which guarantees orthogonality) and using the decomposition \eqref{eq:ranktwo}.

Unless a coupling kernel has some known structure or decomposition we expect it to have a countably infinite rank. For example, a previously used coupling kernel that generalizes the cosine kernel \cite{Sridhar2021,Oscar2023,Gorbonos2024} takes the form
$$J(\theta , \theta') = \mathcal{I}(\theta-\theta') $$
where
$$
\mathcal{I}(\phi) = 
\cos\left [\pi \left ( \frac{d_\mathbb{T}(\phi)} {\pi} \right )^\nu \right ]  \qquad d_\mathbb{T}(\phi) = \min_{m \in \mathbb{Z}} |\phi +2\pi m| \, 
$$
which we will refer to as the \emph{angular distortion} coupling. Here, the neural coupling, $J(\theta,\theta')$, depends only on the angular difference, $d_\mathbb{T}(\theta-\theta')$
of $\theta$ and $\theta'$ which is the smallest angle subtended by an arc between the points $\theta$ and $\theta'$ on the unit circle. For $\nu=1$ this reduces to the cosine coupling and for $\nu <1$ stretches the perceived angular difference between two targets. Following the previous authors, we note this when $\nu \ne 1$ the scaling is \emph{non-Euclidean} - that is angular differences are distorted. They also refer to the case where $\nu <1$ ($\nu>1$) as \emph{elliptic} (\emph{hyperbolic}) as proximal angles are stretched (shrunk). 

In Appendix \ref{app:distort}, we examine this coupling for $\nu=\frac 12$, which is a typical value used in previous studies \cite{Sridhar2021,Oscar2023,Gorbonos2024}. Graphing a $2M+1$ dimensional approximation to the kernel and computing its relative $L^2$ error, the qualitative convergence appears good but the pointwise convergence at the peak is slow due to the cusp: relative error for modes up to order 5 [$\mathbb{E}(5)$] is about 4.4\%.

\subsubsection{A note on computation complexity}

From the viewpoint of neural geometry, naively the computation of the Hamiltonian \eqref{eq:H} scales as the square of the number of neurons, $\bigoh(N^2)$. The spectral decomposition, \eqref{eq:spec_decomp}, allows us to rewrite this sum as
\begin{align*}
H(\vec{n}) &= 
- \sum_{k=1}^K \sum_{\ell=1}^K n_k n_\ell J(\thetav_k,\thetav_\ell)
= 
- \frac{1}{N^2}
\sum_{\summ=1}^N \sum_{\sumn=1}^N \sigma_\summ \sigma_\sumn J(\thetav_\summ,\thetav_\sumn)
\ , \\
&= - \sum_{\lambda_m \in \Lambda} \lambda_m 
\left [\sum_{k=1}^K n_k q_m(\thetav_k) \right ]^2 
=  - \frac{1}{N^2}  \sum_{\lambda_m \in \Lambda} \lambda_m \left [
\sum_{\suml=1}^N \sigma_\suml q_m(\thetav_\suml) \right ]^2 \ .
\end{align*}
Examining this sum,  one can see that the all-to-all coupling may be replaced by computing the projection onto each of the spectral modes, squaring the amplitude of the projection and summing over the modes. From a neuronal computation perspective this reduces the complexity of the calculation from $\bigoh(N^2)$ to $\bigoh(D \cdot N)$ where $D$ is the rank of $J$. In the continuum limit this reduces the complexity from $\bigoh(K^2)$ to $\bigoh(D \cdot K)$ where $K$ is the number of targets.

The desirability of rank two reductions is now evident; the all-to-all coupling for systems with thousands to hundreds of thousands of neurons this seems biologically infeasible. For low rank couplings the number of neural connections only scales as a small multiple of the number of neurons, a much more plausible hypothesis.

\subsection{Suppression of Peripheral Targets}
\begin{figure}
    \centering
    \includegraphics[width=\linewidth]{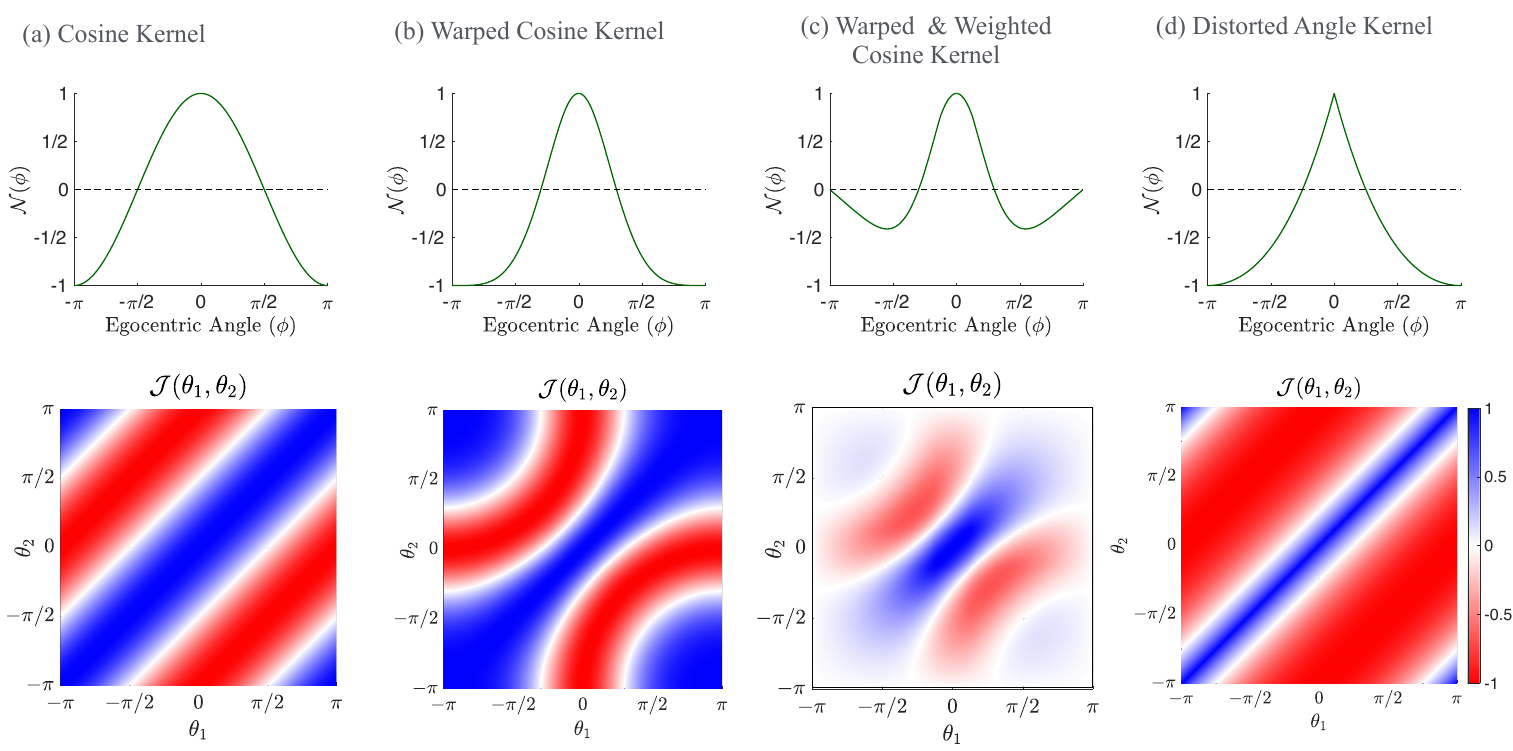}
    \caption{Neuronal coupling for various coupling kernels. Upper Row: The neuronal coupling $\mathcal{N}(\phi)$ between a target directly in front of the observer and a target at angle $\phi$. The neuronal enhancement at small angles \& suppression at large angles is clearly visible. Lower Row: The neuronal coupling $\mathcal{J} (\theta_1 , \theta_2)$ between targets at angles $(\theta_1, \theta_2)$. Columns: (a) Cosine Kernel. Coupling depends solely on the minimal angle between the targets and the division between enhancement and suppression occurs when targets are orthogonal. (b) Warped Cosine Kernel. The peripheral visual field is compressed around the observer's bearing (Eqn. \eqref{eq:linearwarp} with $a=\pi/8$ and $b=\pi$) and then a cosine kernel is applied. The result is suppression occurs at smaller angles, esp. in the forward visual field. (c) Warped and Weighted Cosine Kernel. Same as (b) but now with a weighting corresponding to neuronal density. The coupling is much weaker in the periphery where there are fewer neurons. (d) The Distorted Angle Kernel proposed by \cite{Sridhar2021,Gorbonos2024}. Here, the coupling depends solely on the minimal angular difference between the two angles; small angular differences are amplified.}
    \label{fig:MH}
\end{figure}

It has been well documented that when an observer focuses on a central target, visual input in the periphery can be inhibited \cite{Muller2005,Reynolds2009}. This motivates our choice of a neuronal coupling kernel. Specifically, when an observer concentrates on a central target, three features emerge in the attentional distribution: (i) a peak of attentional enhancement for targets that are adjacent to the central core, (ii) a trough of suppression that enhances concentration on the central target, and (iii) a far peripheral region where the suppression fades due to the sparsity of neurons in the fringe of the visual cortex. This suggests a \emph{Mexican hat distribution} of the attentional field \cite{Muller2005}; an example of this distribution is shown in Figure \ref{fig:MH}(c).

To investigate how choices of neuronal coupling kernels reproduce this phenomena, consider a central target at $\theta=0$ and a second target at $\theta=\thetaobs$. The neuronal coupling in the unwarped egocentric coordinate accounting for the variation in neural weighting is given by
\begin{equation}
 \label{eq:neuronal_coupling}   
\coupling(\thetaobs) = J(0,\thetaobs) \times \frac{\weight(\thetaobs)}{\weight(0)} \ ,
\end{equation}
where we have normalized the coupling so that $\coupling(0)=1$ (recall $J(0,0)=1$). $\coupling(\thetaobs)$ measures the neural enhancement/suppression of peripheral targets and should correlate with the experimentally observed Mexican hat distribution \cite{Muller2005}.
We graph $\coupling(\thetaobs)$ for the various models we consider in Figure \ref{fig:MH}.
We also graph the weighted neuronal coupling kernel
$$\mathcal{J}(\theta_1,\theta_2) = \frac{\weight(\theta_1)}{\weight(0)} \times  J (\theta_1,\theta_2) \times \frac{\weight(\theta_2)}{\weight(0)}$$
which reflects the contribution to the total energy (Hamiltonian) of a pair of targets whose location in the unwarped egocentric coordinates are $(\theta_1,\theta_2)$ accounting for their neuronal weights (again normalized so $\mathcal{J}(0,0) =1 $). The conclusion from these graphs is (i) warping or elliptic distortion compress the set of angles where peripheral enhancement is observed and (ii) a drop in neuronal suppression in the far periphery can be obtained by incorporating neuronal density.  

\subsection{The Neural Consensus}
For a coupling kernel of dimension $D$, the Glauber dynamics can be reduced to a $D$ dimensional system of ODEs. This reduction is known for the cosine kernel \cite{Sridhar2021,Gorbonos2024} but in Appendix \ref{app:spectralGlauber}, we derive a method that applies to any finite-dimensional kernel. Here, we will briefly sketch the process using the cosine kernel as an illustration.

Define a complex vector sum over the fraction of excited neurons in each group,
$$\gamma(\tau)\equiv R(\tau) e^{i \Theta (\tau)} = \sum_{k=1}^K n_k e^{i \thetav_k}  \quad \Rightarrow \quad  R(\tau) = |\gamma (\tau)| \ , \quad \Theta (\tau) = \textrm{Arg}[\gamma(\tau)].   
$$
The sum of the $n_k$'s is unity, so the amplitude of $\gamma$ satisfies $0\le R \le 1$ and measures the \emph{coherence} of the weighted average of the directional unit vectors $e^{i \thetav_k}$.
$\gamma$ is in the warped coordinate system where the neural computations are performed with the neural angle $\Theta\in[-\pi,\pi]$; a value of zero indicates the consensus is aligned with the observer's orientation and a positive (negative) value indicates the left (right) of the observer's heading. If $\gamma$ is on the negative real axis, one can choose either
$\Theta=+\pi$ or $-\pi$; as will become clear below, navigationally this choice is moot.

We now posit that the neural consensus should depend only on the spectral amplitudes of $\gamma$ at equilibrium, as these amplitudes are computed independently of target configuration. Let $\qv(\phi)$ be a measure (density) for active neurons, here taken to be $\qv (\phi) \equiv  \sum_{k=1}^K n_k \delta(\phi-\thetav_k)$ as defined in Appendix \ref{app:coupling}. For the cosine neural coupling \eqref{eq:cosineneural} where $Q_+(\phi)=\cos(\phi)$ and $Q_-(\phi)= \sin(\phi)$ (rank two),
$\gamma$ reduces to 
\begin{align*}
\gamma &= \gamma_+ + i \gamma_- =  \langle \qv(\phi), Q_+(\phi) \rangle + i \langle \qv(\phi), Q_+(\phi) \rangle \\
&=\sum_{k=1}^K n_k [Q_+(\thetav_k) + iQ_-(\thetav_k) ] =  \sum_{k=1}^K n_k [\cos(\thetav_k) + i\sin(\thetav_k) ] \\
&= \sum_{k=1}^K n_k e^{i \thetav_k} 
\end{align*}
which is identical to the aforementioned definition but allows generalization to any rank two neural consensus model. 

Using a separation of timescales assumption, the evolution equation \eqref{eq:gammaODE_ab} simplifies significantly for $\gamma$ with the cosine neural coupling,
\begin{equation}
\label{eq:gammaODE}
    \frac{d\gamma}{d\tau} = \sum_{k=1}^K \frac{\rho_k e^{i\thetav_k}}
{1+\exp{\left(-2\beta \Real [\gamma e^{-i \thetav_k}]\right)} }- \gamma
\end{equation}
which is a single complex ODE for $\gamma$.
We've used the fact that  
$${ (\Delta E)_{k} = 
-2 \left [ \gamma_+ Q_+(\thetav_k)
+\gamma_-Q_-(\thetav_k) \right ]
= -2 \left [ R \cos (\Theta) \cos(\thetav_k)
+ R \sin(\Theta) \sin(\thetav_k) \right ] =
- 2 R \cos(\Theta-\thetav_k)= -2\Real [\gamma e^{-i \thetav_k}] }$$
to close the system (see Appendix \ref{app:cosine_kernel}).
This is a gradient flow, as shown in Appendix \ref{app:gamstability}. Note that other coupling kernels, including the angular distortion discussed earlier, are more difficult to express precisely in terms of $\gamma$ and may require $\Delta E$ to be expressed in terms of the target system $n_k$ to remain gradient flow.

The neural amplitudes evolve to an equilibrium $\vec{n}^*$ on the fast timescale $\tau$; this allows us to define the \emph{neural consensus}
$$\gamma^* = R^*e^{i \Theta^*} = \gamma_+^* + i \gamma_-^* =  \sum_{k=1}^K n_k^* e^{i \thetav_k}  $$
which satisfies the implicit equation 
\begin{equation}
\label{eq:Gammastar}
\gamma^* = \sum_{k=1}^K \frac{\rho_k e^{i\thetav_k}}
{1+\exp{\left( -2\beta \Real [\gamma^* e^{-i \thetav_k}] \right)} }.
\end{equation}
The neural consensus direction $\thetabar$ and coherence $\Rbar$ can now be used in a navigational model.

\subsection{The navigational model}
\label{sec:navigation_model}

The study of agent-based models has produced a broad variety of models for navigation \cite{BasRom2020,VicZaf2012,GaoCou2025}. Most assume that sensory input is processed and results in a modification of an individual's position, orientation, and/or momentum related to attraction or avoidance of objects. We will restrict ourselves here to a simple model where an individual's speed is assumed to be constant, and their bearing evolves in response to neural consensus. This is in contrast to previous work where an individual's velocity is chosen to be proportional to the neural consensus vector \cite{Sridhar2021,Gorbonos2024}.

In the allocentric (laboratory) frame, we define a configuration space for a walker consisting of a position on the plane, $\vec{x}= \langle x,y \rangle$, and a bearing measured as a polar angle $\varphi$ (in mathspeak $ (\vec{x},\varphi)  \in \mathbb{R}^2 \times S^1$).
An individual's trajectory in this space is given by $\vec{x} = \vec{X}(t)$ and its bearing by $\varphi=\bearing(t)$, and both are specified by a set of differential equations
\begin{subequations}
\label{eq:walkerDE}
\begin{align}
   \frac{d\vec{X}}{dt}  &= V \hat{\bearing} \ , \qquad \hat{\bearing} = \langle \cos \bearing , \sin \bearing \rangle  \ ,\label{eq:walkerX} \\
    \frac{d\bearing}{dt}&= \Krate\Rbar \sin(\thetabar/2) . \label{eq:walkerPhi}
\end{align}
\end{subequations}
where $V\in\mathbb{R}$ is the walker's speed and $\Krate$ is a physical constant related to turning speed. In particular, Eqn. \eqref{eq:walkerPhi} describes an impetus for the animal to change its direction based solely on the neural consensus $\gamma^* = R^*e^{i \Theta^*}$. It is a completely egocentric model of observer reaction to stimuli and neural decision-making.
Since $\thetabar\in[-\pi,\pi]$, taking the argument of the sine function to be $\thetabar/2$ ensures that the walker turns towards the consensus angle at a rate that increases monotonically to a maximum at the edge of the visual field. In the extreme case where $\thetabar= \pm \pi$, the turning rate is discontinuous. We choose $\frac{d\bearing}{dt}=0$ at this point, but exclude it from the equilibria sets described below.

This model is deterministic and can be used to determine the existence and stability of steady-state bearings at different points in space (see Appendix \ref{app:navstability}). In section \ref{sec:Langevin} below, we extend this model to include noise as a function of the coherence $\Rbar$.

\subsection{Adding noise: The Langevin Model}
\label{sec:Langevin}
It is often desirable to model organismal movement as a random walk. Since neural processes occur quickly compared to physical turning, we assume noise in $\gamma$ averages out on the timescales relevant to navigation. Therefore, we only consider adding model noise to Eqn. \eqref{eq:walkerPhi} and propose the Langevin equation
\begin{equation}
    d \Phi  = \Krate (\Rbar)^p\sin(\thetabar/2) \, dt  + \sigma(1-\Rbar)^q\cos(\thetabar/2)\ dW,\label{eq:Langevin}
\end{equation}
where $p,q\geq 0$ and $\sigma>0$ is the standard deviation of the Wiener process $W$.  

This formulation imagines movement as a combination of directed motion toward targets on the one hand and an unbiased random walk on the other, with $\thetabar=\Rbar=0$ in the visual absence of targets. Behavior between the two is mediated by the coherence value $\Rbar$ while $p$ and $q$ adjust the rate. In particular, $\sigma$ may need to be relatively large to reproduce desired random walk behavior when $\Rbar=0$; choosing $q>1$ allows this to scale so that noise does not dominate for nominal values of $\Rbar$. Similarly, we expect that $0<p\leq 1$.

Multiplying by cosine in the noise term means that when facing a consensus direction ($\thetabar=0$), neural mediated drift is zero and the noise is full: the walker explores around its on-course heading. When facing away from the consensus direction, noise goes to zero and the corrective neural impulse is clean and dominates. One additional note about Eqn. \eqref{eq:Langevin} is that in the single-target case, $\Rbar=1$ and there is no noise. One could add an additional (small) Gaussian noise term calibrated to this case and adjust $\sigma$ accordingly.





\section{Results}

\begin{figure}[t]
    \centering
    \includegraphics[width=\linewidth]{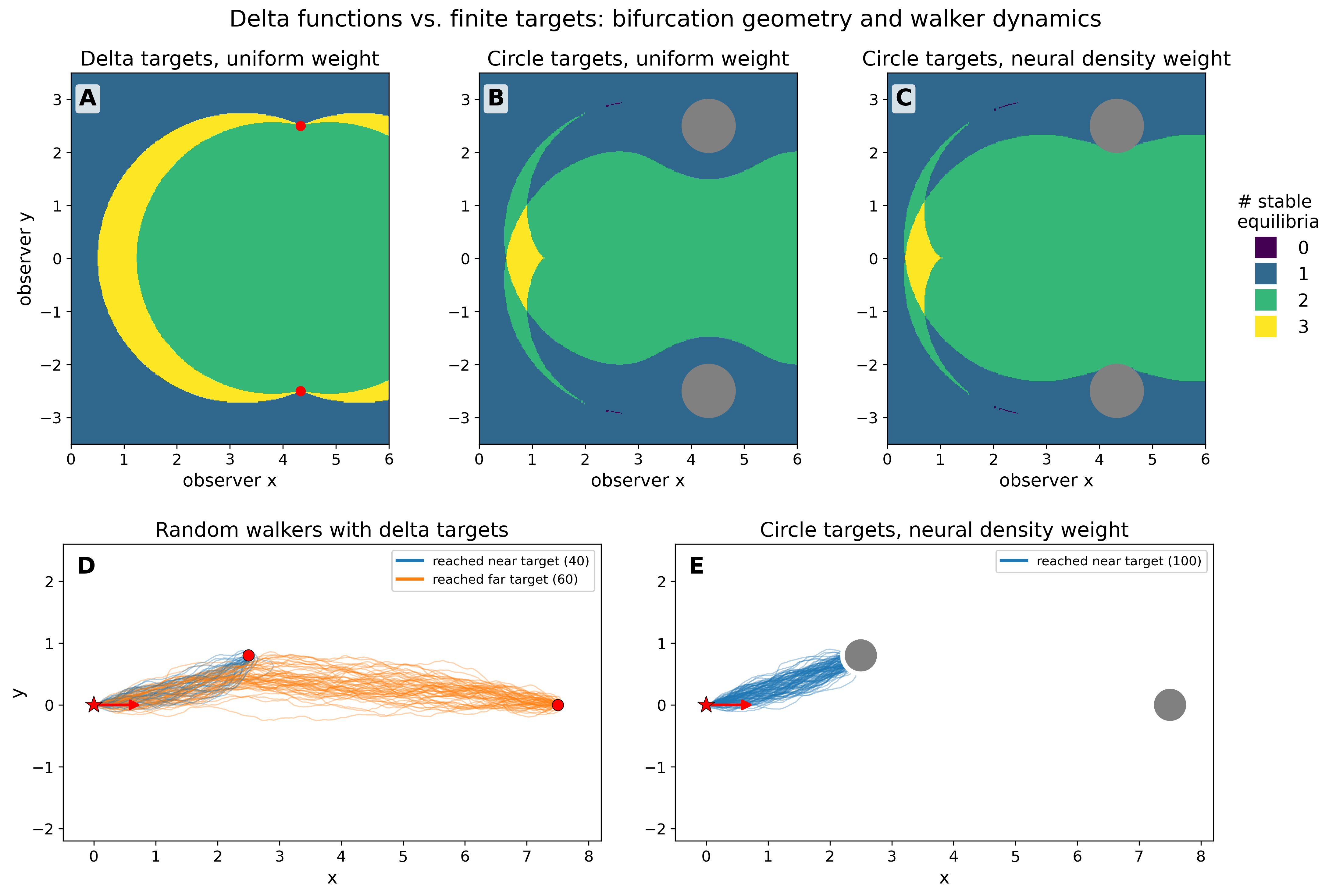}
    \caption{Bifurcation diagrams showing the number of self-consistent stable equilibria in $\theta$ as a function of spatial position for (A) delta function targets, and (B)-(C) targets of radius 0.5. All targets are centered at $(4.33, \pm 2.5)$ and a linear-cutoff function was used for the neural density with $a=\pi/8$ and $b=\pi$. In panel (B), targets were weighted uniformly by angular position (unweighted case) while in (C), the neural density function was used for target weighting. Panels (D) and (E) show 100 walkers starting at the origin facing along the positive $x-$axis with additive angular, Gaussian noise with $\sigma=0.5$ radians. Panel (D) uses delta function targets and (E) uses circular targets of radius 0.25. Walkers were considered to have found the target if they come with 0.1 of it. Model parameterization for panel (D) follows that of (A) while parameterization for (E) follows that of (C); using a uniform weight instead produces qualitatively equivalent results to (E) and so is not shown.}
    \label{fig:delta_vs_circles}
\end{figure}

Using the stability criteria described in Appendix \ref{app:ndstability}, we can visualize the dynamics of our framework by counting the number of \emph{self-consistent} (SC), locally asymptotically stable equilibrium points - that is, asymptotically stable equilibrium points for which $\gamma^*\in\mathbb{R}$ and $d\theta/dt=0$ in Eqn. \eqref{eq:walkerPhi} - as a function of location in $(x,y)$-space. The geometry of the resulting bifurcation diagram is then a function of target placement, size, and type and differs notably in the case of delta-function targets versus targets of finite extent (Fig. \ref{fig:delta_vs_circles}). 

In the case of two delta-function targets, the first bifurcation a walker would encounter while moving toward the targets along $y=0$ is characterized by a crescent-shaped region connecting both targets. For high enough values of $\beta$, there is a region of tri-stability (as noted in \cite{Sridhar2021,Gorbonos2024}) before a larger region of bi-stability. See Fig. \ref{fig:delta_vs_circles}A for details. For targets of finite extent, the geometry is more complex: the multi-stable region is typically characterized by a bulbous region of bistability with horns that extend out toward either target, and tri-stability occurs within a more contained region where the horns connect to the main bulk of the shape (see Fig. \ref{fig:delta_vs_circles}B-C). The bifurcations and dynamics corresponding to each of these regions will be explained shortly.

\begin{figure}[t]
    \centering
    \includegraphics[width=\linewidth]{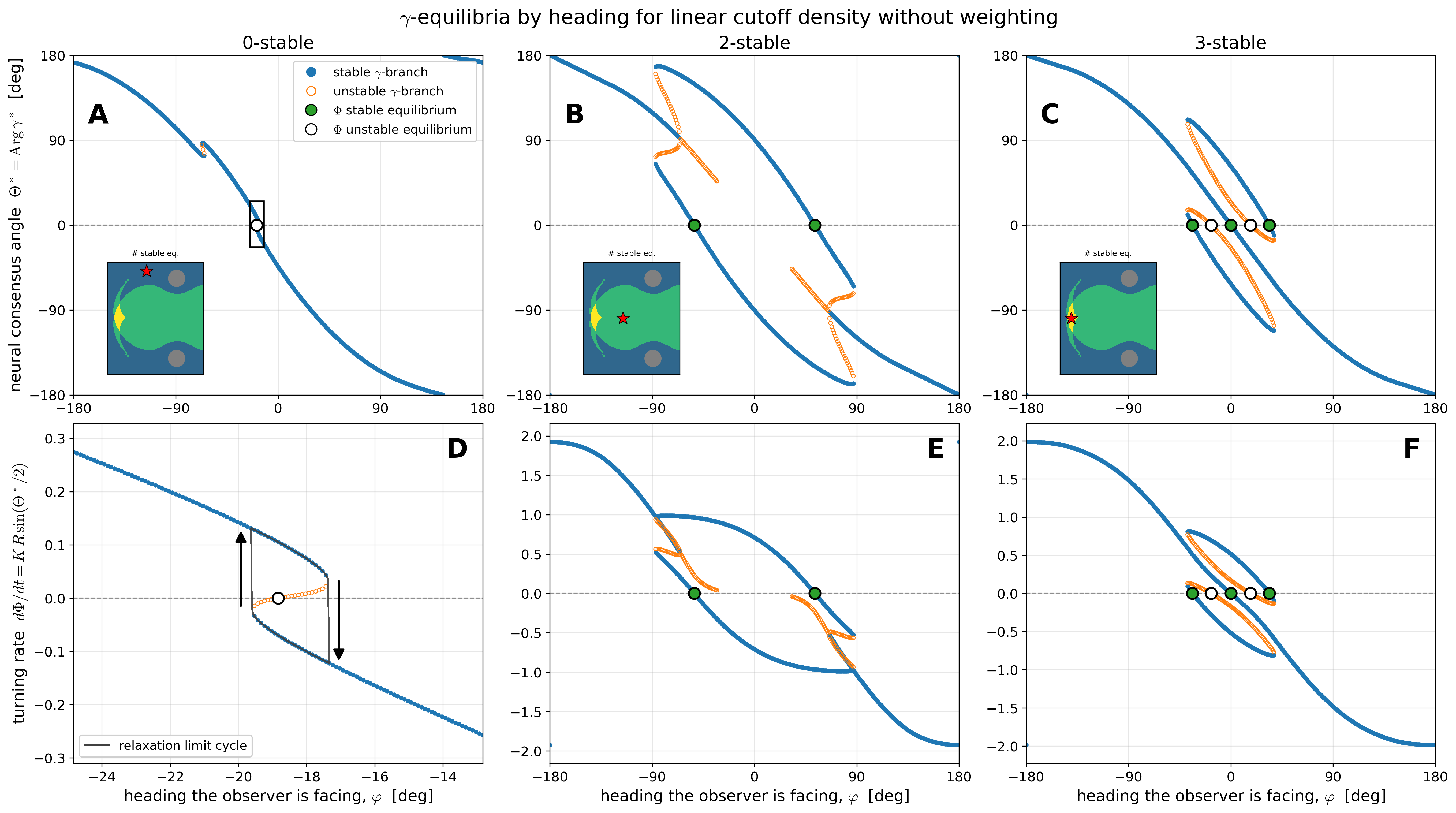}
    \caption{Bifurcation diagrams at fixed $(x,y)$ coordinates while allocentric observer heading varies. (A) $\Arg(\gamma)$ as a function of observer heading in a location with no stable self-consistent equilibria. Locations where the graph crosses the x-axis are also equilibria in $\theta$. (D) Zoom-in of square area in (A) showing hysteresis loop. The next two columns are $\Arg(\gamma)$ as a function of observer heading in a location with two (B) and three (C) stable self-consistent equilibria, with (E) and (F) showing $d\Phi/dt$ as a function of heading for these two cases respectively. Stable and unstable self-consistent equilibria are marked. Each case has two circular targets of radius 0.5 at $(4.33, \pm 2.5)$, a linear cutoff neural density with $a=0$, $b=\pi$, and no neural weighting. $\beta=10$ and $\Krate=2$.}
    \label{fig:theta_bifurcations}
\end{figure}

For all cases of two targets, regions typically contain 1-3 SC stable equilibrium points corresponding to each target and a consensus direction between them. The attracting state of the system for a given spatial configuration $(x,y,\varphi)$ is a function of the initial neural angle and coherence value. In some cases (Fig. \ref{fig:delta_vs_circles}B-C), regions that contain zero SC stable equilibrium points also appear. These regions correspond to hysteresis relaxation loops (see Fig. \ref{fig:theta_bifurcations}A and D).

In terms of practical effects on a random walker making decisions while on the move, there is a profound difference in decision-making behavior with regard to delta-function targets versus even very small targets of non-zero radius. A walker with moderate angular noise that approaches a nearby $\delta$-function target with a much further $\delta$-function target near the center of its view will very often choose the further of the two. This occurs regardless of any weighting of targets by egocentric position $\theta$ (Fig. \ref{fig:delta_vs_circles}D) because $\delta$-targets do not change neural group size $|G|$ according to distance: their extent is always infinitely small, and so their group size remains fixed. However, targets of finite extent grow in apparent size as they get closer, which in turn increases $|G|$ and makes them more attractive. For any reasonable amount of noise, observers are therefore always drawn to the nearer of the two targets rather than the one further behind it (Fig. \ref{fig:delta_vs_circles}E).

For the remainder of these results, we will consider only circular targets with positive radius since this is a more biologically realistic scenario. Parameterization of the model requires a configuration in $(x,y,\varphi)$-space for the observer's position and heading. Then, since $d\gamma/d\tau$ is a gradient flow, we are guaranteed that any initial condition $\gamma_0$ will asymptotically approach a stable equilibrium as $\tau\rightarrow\infty$. However, the stable equilibrium it approaches depends upon $\gamma_0$, and it will often have $\Arg(\gamma^*)=\Theta^*\neq0$. For any nonzero $\Theta^*$, Eqn. \eqref{eq:walkerPhi} for $d\Phi/dt$ turns the observer thereby changing its heading and moving the egocentric angular locations of the targets. This procedure repeats, usually as the observer is also changing its spatial position. If instead $\Theta^*=0$, then $d\Phi/dt=0$ as well and (discounting noise) the observer will maintain its heading. We refer to this as a self-consistent equilibrium.

As a result, for any position in $(x,y)$-space we can take observer bearing $\varphi$ as a bifurcation parameter and observe the equilibrium structure of $\Theta^*$ with particular attention to points that cross the $\Theta^*=0$ axis (see Fig. \ref{fig:theta_bifurcations}A-C). The structure of these bifurcation diagrams hint at how SC-equilibrium directions appear and disappear as one moves around in $(x,y)$-space; for example, a walker moving left to right along the $x$-axis will first have one SC-equilibrium point (at $\varphi=0$) and then through two saddle-node bifurcations, it will end up in a tri-stable regime with basins of attraction separated by unstable SC-equilibria in the forward direction and a branch cut behind (Fig. \ref{fig:theta_bifurcations}C).

Since Eqn. \eqref{eq:walkerPhi} is strictly increasing and keeps the sign of $\Theta^*$, the dynamics of $d\Phi/dt$ with respect to $\Theta^*$ are geometrically similar to the bifurcation diagram of $\Theta^*$, with $\Theta^*>0$ corresponding to increases in $\Phi$ and $\Theta^*<0$ corresponding to decreases in $\Phi$ (see Fig. \ref{fig:theta_bifurcations}D-F). In some cases, this can mean the observer turns through an otherwise SC-equilibrium direction to settle on a very different direction. For example, in Fig. \ref{fig:theta_bifurcations}B and E, an observer that starts out facing at approximately $\varphi=-80^\circ$ on the top blue branch instead of the bottom one would begin facing slightly to the right of the $\varphi$-stable equilibrium corresponding to the right target. But with $d\Phi/dt>>0$, it would turn to the left following the top blue branch down until it crosses the $d\varphi/dt$-axis at the location corresponding to the left target equilibrium. Note that which stable branch an observer follows is ultimately a function of its initial neural state, which means both $\Theta_0$ \textit{and} coherence $R_0$: a less committed observer (low $R$) starting from the same spatial configuration and neural angle may end up at a different target than one that is already highly committed to a particular direction.

\begin{figure}[ht!]
    \centering
    \includegraphics[width=\linewidth]{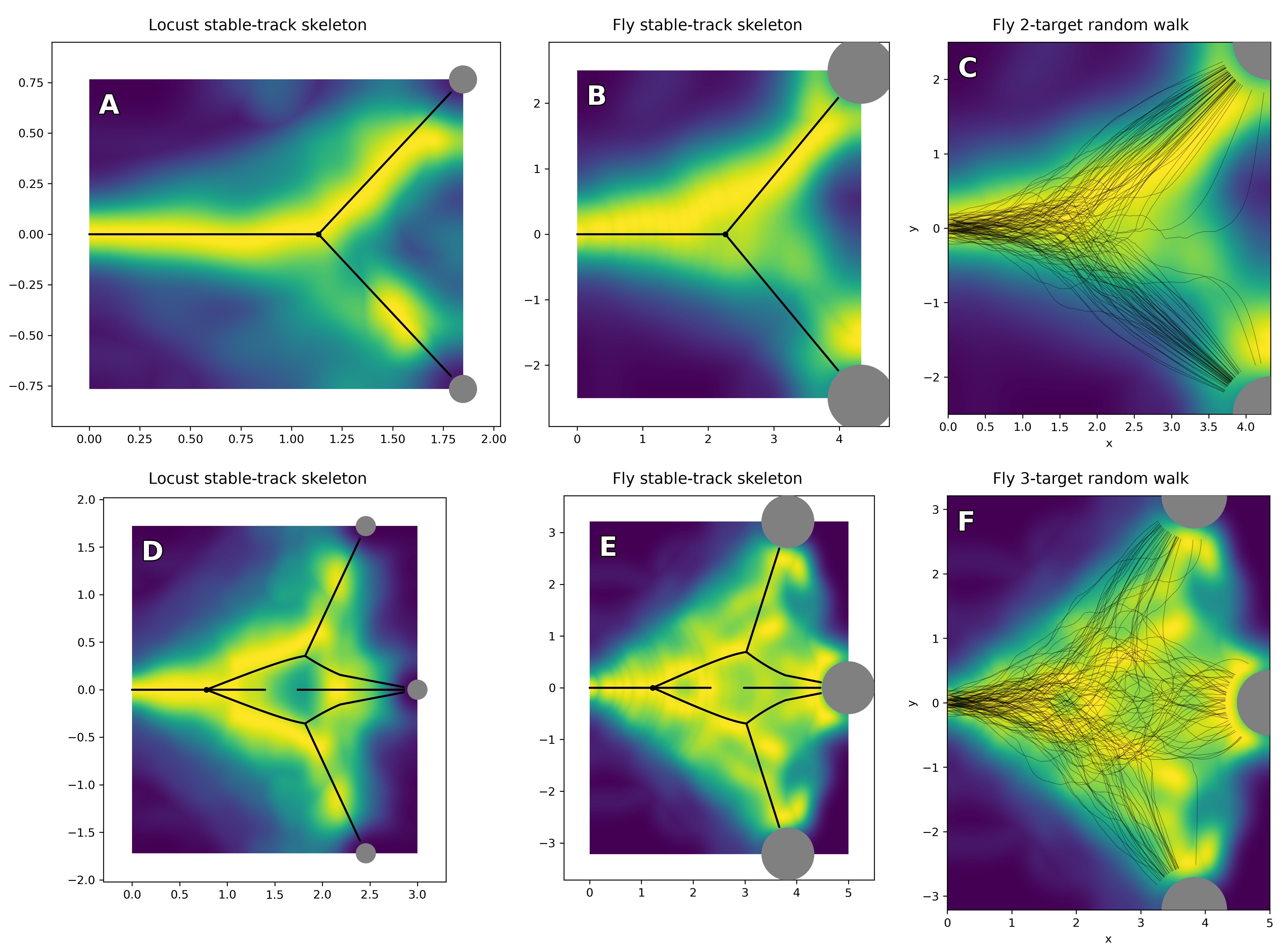}
    \caption{Diagrams showing stable equilibrium directions for the (A) two-target locust experiment, (B) two-target fly experiment, (D) three-target locust experiment, and (E) three-target fly experiment over histogram for all animal track data from Sridhar et al. \cite{Sridhar2021}. Each black track was generated from numerical calculations corresponding to the underlying bifurcation structure; note in (D) and (E) that the central branch loses and then regains stability. Attempts were made to qualitatively fit the model with Eqn. \eqref{eq:Langevin} to the histogram data including the ratio of walkers finding each target. (C) and (F) show the results for the 2-target and 3-target fly experiment respectively using 1500 tracks. A linear cutoff neural density was used with $a=0.65\pi$ and $b=0.92\pi$ and a linear cutoff weighting with $a=0.2\pi$ and $b=0.8\pi$. $\beta$ was set to 20 in the two target case and 30 in the three target case to match the noise profile in \cite{Sridhar2021}. $\Krate=2$, $p=3$, $q=2$, and $\sigma=4$ in the Langevin equation with a walker speed of $0.3$ m/s. Walkers found the target if they were within 0.2 m. Parameterization for (B) and (E) were identical; for (A) and (D) parameters were chosen to best represent locust track data but the target ratio was unrecoverable. Parameters were $\Krate=6$, the same $\beta$ values as above, $a=0.5$ and $b=0.9$ for the linear cutoff neural density and $a=0.1$ and $b=0.8$ for the linear cutoff weighting.}
    \label{fig:walkers}
\end{figure}

Off-axis dynamics (the ``horns'') are illustrated in Fig. \ref{fig:horn_dynamics} in Appendix \ref{app:model_details}. Moving from left to right through this region, dynamics go from that of a single stable SC-equilibrium, corresponding to a consensus direction between targets, to bistability between the near-target and the consensus direction through a backward fold bifurcation. The horn region then ends in a saddle-node bifurcation that eliminates the consensus direction. The single stable SC-equilibrium region past the horn arc then corresponds to the near target while entering the main bi-stable region corresponds to another fold bifurcation that gives rise to a stable SC-equilibrium corresponding to the far target. It is important to keep in mind that in terms of stable SC-equilibria, basins of attraction in $\varphi$ (with $\gamma_0$ chosen from a nearby $\gamma^*$) are not always cleanly separated by unstable equilibria. Numerically, we have seen unstable equilibria, $\gamma$-folds, branch cuts (due to a target being directly behind an observer with $360^\circ$ view), and disappearing targets (due to a blind spot; e.g. $b<\pi$ for the linear cutoff neural density function) all form the boundary of a basin of attraction in different parameterization regimes.


An organism that is traversing the landscape while making decisions about which target to explore will follow predictable paths through the bifurcation structure we have previously described based on assumptions of continuity that apply both to its neural states and to its position and heading as the observer moves through space. 
To illustrate these paths, we adopt target geometries from \cite{Sridhar2021} corresponding to their lab experiments with both fruit flies and locusts and trace out the stable paths a walker will take when starting from the origin facing $\varphi=0$ (Fig. \ref{fig:walkers}A-B for two targets and Fig. \ref{fig:walkers}D-E for three targets). Histograms for the unfiltered \cite{Sridhar2021} experimental data set showing the paths real organisms took to the targets are included. In all cases, the bifurcation structure begins with a single consensus path toward the middle of the targets. In the two-target case, a walker can continue to follow the center path through any tri-stable region (i.e. Fig. \ref{fig:theta_bifurcations}C) or switch due to $\theta$-noise; however, at the end of the tri-stable region, the center stable path disappears and only two stable paths remain: one leading to each target. This results in the ``Y'' structure seen in previous works \cite{Sridhar2021,Gorbonos2024}.

\begin{figure}[t!]
    \centering
    \includegraphics[width=\linewidth]{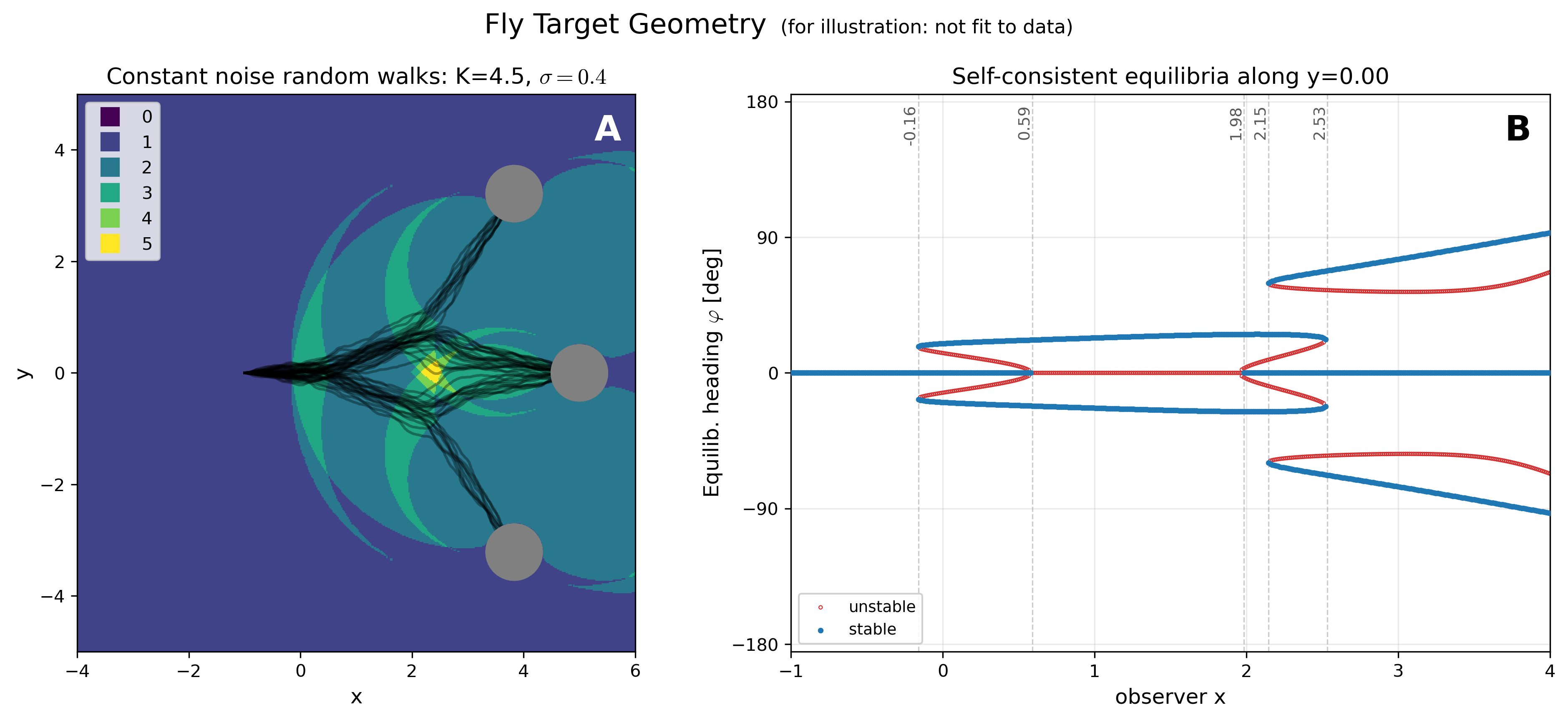}
    \caption{Bifurcation diagrams showing the number of self-consistent stable equilibrium points for 3 targets according to the target size and geometry used in \cite{Sridhar2021} (targets are located at $(5,0)$ and $(3.83,\pm 3.214)$). Model dynamics were not fit to the experimental data; instead, an identical linear cutoff function was used for both $\mu(\theta)$ and $\omega(\theta)$ with $a=0.25\pi$, $b=0.9\pi$, and $\beta=15$. In (A), tracks for 50 walkers starting at $(-1,0)$ with heading $\varphi=0$ are plotted on top. To visualize the center bias, constant $\theta$-noise was used with $\sigma=0.4$ and $\kappa=4.5$ instead of Eqn. \eqref{eq:Langevin}. (B) Bifurcation diagram along $y=0$ demonstrating how self-consistent headings gain and lose stability.}
    \label{fig:3target_branch}
\end{figure}

The bifurcation picture is somewhat different in the three-target case, as shown in detail in Fig. \ref{fig:3target_branch}. The first bifurcation gives rise to a region of tri-stability between the center target and two consensus directions between the center and outer targets. Angular noise can put a walker on any of these, but then the center stable direction disappears in a subcritical pitchfork bifurcation (Fig. \ref{fig:3target_branch}B, and also the missing slice at $x\approx 1.5$ in Fig. \ref{fig:walkers}D and $x\approx 2.5$ in Fig. \ref{fig:walkers}E). This recovers the behavior seen in Sridhar \textit{et al.} \cite{Sridhar2021} for the $\delta$-function case. 

However, with targets of finite size something different then happens. Before the consensus directions go unstable, the center direction becomes stable again in a second (backwards) subcritical pitchfork bifurcation (Fig. \ref{fig:3target_branch}B and $x\approx 1.7$ in Fig. \ref{fig:walkers}D, $x\approx 2.9$ in Fig. \ref{fig:walkers}E). The consensus directions remain stable as this happens but then disappear in saddle node bifurcations that are a mirror image to the structure that gave rise to them. Slightly before, stable directions to the outer targets appear through separate saddle node bifurcations (Fig. \ref{fig:3target_branch}B). This results in a tiny window on the $x$-axis where there are actually five concurrently stable SC-equilibrium points.

The practical effect of these dynamics is that without noise or a navigational model with directional inertia, walkers tend to choose the central target (see Fig. \ref{fig:3target_branch}A). Tracks begin by following a central consensus path before branching off to the left and right along consensus directions. They then come back together toward the central target ($100\%$ of the time with no noise and a large proportion of the time with low, constant noise; see Fig. \ref{fig:3target_branch}A). 

The navigational Langevin model that we propose in Eqn. \ref{eq:Langevin} is capable of overcoming this behavior by balancing a significant amount of noise when the walker is uncommitted and $R^*$ is relatively low (enough noise that when no targets are visible and $R^*=\Theta^*=0$, dynamics are a pure Wiener process in $\theta$ with standard deviation $\sigma=4$, thus approximating a standard Brownian search behavior) with target homing for a committed walker that has $R^*$ close to 1. Weighting targets that are closer to the egocentric center of vision higher while suppressing those in the periphery also helps. Adjusting parameters in the neural density function can then align the locations of the bifurcations to match with experimental data (see Figs. \ref{fig:sweep_a_uniform_weight}-\ref{fig:sweep_b_uniform_weight} in Appendix \ref{app:model_details}). The bias toward the central target is in some respects a consequence of geometry since there are no targets to the left and right of the outer targets to balance the bifurcation space; see Fig. \ref{fig:9target} in Appendix \ref{app:model_details} for a bifurcation diagram with a complete circle of targets.

In tuning our model to match experimental data, we noted that tracking data from \cite{Sridhar2021} suggests that fruit flies chose the center target approximately half the time and the outer targets 25\% of the time each. This ratio is precisely what one expects of an organism that chooses one of the two consensus directions and then either the central target or an outer one. With manual adjustment of parameters, our modeling structure was able to reproduce this ratio for the three-target case while matching the approximate locations of the bifurcations (see Fig. \ref{fig:walkers}C and \ref{fig:walkers}F). However, the locust track data from \cite{Sridhar2021} suggests that locusts chose the outer targets 71\% of the time and the central target only 29\% of the time, implying they are actually biased toward the outer targets. Our navigational model was unable to reproduce this ratio. However, locusts exhibit jerky, stop-and-go motion with jumps that may be a poor match for both our continuous Langevin navigational model and our assumption that the neural timescale and physical timescale are fully separated. Additionally, as proposed in \cite{Sridhar2021}, bias toward outer targets may simply be a biological feature of decision making for some organisms that is essential to correctly predicting behavior.

Since targets with positive physical area appear larger when an observer gets closer versus delta-functions, it can be desirable to quantify the extent of the basins of attraction for stable SC-equilibria when in multi-stable regions of bifurcation space. However, the extent of these basins is hard to characterize in general. SC-equilibria are always real-valued (and so we can choose $\Theta_0=0$ for any given $\varphi$), but the asymptotic dynamics can still be a function of $R_0$. For example, an observer with a strong coherence may pick the target it is closer to facing while an observer with the same heading but a weaker coherence may continue walking toward a consensus direction. There are also scenarios within a bistable regime where for a given $\varphi$, perhaps pointed somewhat away from a nearby target, the observer is attracted toward the near target for most values of $R_0$ while for high $R_0$, the further target attracts. Additionally, basin edges are not simply characterized by unstable equilibria: there can be folds and blind spots. The former of these in particular means that how wide a given basin is can depend on which side you approach from. One can also ask questions about robustness of any given stable SC-equilibrium with respect to a given noise model.

To simplify this situation while still visualizing basin extent, we suggest using an approach that targets the behavior of an uncommitted observer with $R_0=0.15$ and $\Theta_0=0$. Then, since $d\gamma/d\tau$ is a gradient flow, for almost any configuration $(x,y,\varphi)$ for which there is at least one stable SC-equilibrium this initial condition is guaranteed to asymptotically approach just one of the stable equilibria. Fig. \ref{fig:fly_bifurcation} Appendix \ref{app:model_details} demonstrates a result using this method. However, for a walker that has already been traversing space, it is often much harder for $\theta$-noise alone to initiate a change in decision because $R$ may be quite large, corresponding to a strong coherence in the neural band. In future work, it may be interesting to explore a modeling regime where large, stochastic fluctuations in $R$ are possible, for instance, following a transition from moving to stationary.

\section{Conclusions}

In this paper, we introduced the concepts of a neural mapping (Eqn. \eqref{eq:angle_map}) and neural group size (Eqn. \eqref{eq:group_size}) in order to derive a mean-field model of a neural ring attractor that is fully capable of tunable, spatial decision-making among targets of finite size while utilizing an interaction kernel that is only of rank 2. Neural group size is also capable of applying a foveal effect where targets in front of the observer are given a higher weight than those in the periphery. When deriving a mean-field model for the Glauber dynamics of our formulation, we discuss the role of neural coherence $R$ in addition to the neural consensus direction $\Theta$ and then introduce a Langevin model for navigation that utilizes the coherence value to balance Gaussian random walk searching behavior with attraction to targets.

Applying our framework to scenarios of spatial decision-making, we demonstrate the singular nature of the $\delta$-function case and compare the geometry of decision-making between foveal and non-foveal cases (Fig. \ref{fig:delta_vs_circles}). It is additionally possible to restrict the field-of-view for an observer by choosing a neural density function $\mu(\theta)$ that is supported only on a subset of $[-\pi,\pi]$, thereby implementing a blind-spot behind the animal. Doing so can create regions of bistability between targets in a manner similar to foveal weighting because the observer loses the target behind them, even if it is relatively close. We also demonstrate (Fig. \ref{fig:theta_bifurcations}) that the bifurcation structure in multi-stable regions of the $(x,y,\varphi)$ configuration space is often complex: neural consensus $\gamma^*$ is two dimensional, and so the attracting neural consensus direction $\Theta^*$ is ultimately a function of both $\Theta_0$ and $R_0$ (initial values for the neural band) as well as the spatial configuration of the observer.

Our framework is largely able to reproduce experimental data for two-target cases of spatial decision-making; however, the bifurcation structure for three targets raises many questions and suggests there may be additional biological mechanisms that would allow organisms to conduct a bifurcation cascade on finite-sized targets focused on subdividing potential choices and then choosing between the targets within each subdivision. As we have demonstrated, noise has a role to play, but neural noise was averaged out in the derivation of our mean-field model through the separation of timescales. Unlike flies, locusts exhibit jerky, quick movements that may make the time-scale separation assumption a poor one. Additionally, differential models of turning like the Langevin equation in Eqn. \eqref{eq:Langevin} are best suited for relatively smooth motion such as that seen in flies and less suited to capture the stop-and-go motion of locusts. For locusts in particular, a logical next step is to think about tuning the navigational model with the available field data \cite{Weinburd2024,Gorbonos2024b}. 

Still, re-formulations of model noise or the turning model will not alter the underlying bifurcation structure of three-target decision-making reported here. One idea that was tried and discarded is to use an anti-foveal neural weighting, $\omega(\theta)$, essentially biasing the observer away from targets that are directly in front of it using a function that dips around $\theta=0$. This actually made the problem worse because outer targets are already weaker, and an anti-foveal weighting in egocentric angle $\theta$ suppresses whatever is dead ahead without knowing whether that is the center or an outer target. For the $\delta$-function target case, Sridhar \textit{et al.} \cite{Sridhar2021} proposed an `overlap' function in their supplementary information in order to account for experimental observations that animals treat directionally-similar targets as a single unit, still visiting an attractive but directionally distant target. This function works by discounting a proportion of spins encoding a target if there are other targets in directional proximity. If we re-construe ``other targets in directional proximity'' to refer instead to the spatially-continuous extent of our now finite-sized targets which in turn informs neural group-size, a similar affect of de-biasing the center might be achieved.

Our results suggest that experimentation aimed at the region between two targets may be important for correctly fitting a neural band model, since the bifurcation point where bistability collapses into attraction to only the nearby target is tuneable through neural weighting. We also predict that for a sufficiently off-axis observer approaching two targets of finite extent, there is no region of tri-stability; this stands in contrast to the delta-function case (see Fig. \ref{fig:delta_vs_circles}) and is another prediction that can potentially be evaluated experimentally. Finally, our characterization of basin-of-attractions for an uncommitted observer is a measurement that could be quantifiably compared to data (see Fig. \ref{fig:fly_bifurcation} in Appendix \ref{app:model_details}). Using a model parameterization in which the bifurcation points have previously been fit to tracking data, we might expect that the basins of attraction roughly correspond to the probability that an observer, suddenly released at point $(x,y)$ in space, is attracted to each of the possible self-consistent directions.

\bigskip
\noindent \hrulefill
\bigskip

\noindent \textbf{Data access:} The Python and MATLAB code that was used to conduct simulations and numerically analyze the model will be made available on Github upon acceptance.

\bigskip

\noindent\textbf{Author contributions:} WCS and AJB contributed equally to all aspects of this study.

\bigskip


\noindent\textbf{Funding:} This work was supported by the National Science Foundation under Award No. 2410988 to WCS and the American Institute of Mathematics under the SQuaREs program, funded by the National Science Foundation and the Fry Foundation. WCS would like to acknowledge the Simons Foundation Collaboration Grants for Mathematicians which helped support travel and meetings related to this work, and travel support from the Max Planck Institute for Animal Behavior to visit the Institute. AJB acknowledges the support from the Australian Research Council under the Discovery Projects scheme (Project Number DP260101231).

\bigskip

\noindent\textbf{Acknowlegements:} This work was inspired by WCS's visit to the Max Planck Institute for Animal Behavior, Department of Collective Behavior during his professional development leave, and WCS would particularly like to acknowledge and thank Dr. Iain Couzin and Dr. Dan Grobonos for their scholarly support and guidance, both at the Institute and afterward, without which this work would never have gotten off the ground. WCS and AJB would also like to acknowledge support from the rest of the 2026 SQuaREs team: Drs. Rebecca Everett, Maryann E. Hohn, and Jasper Weinburd.


\vfill
\eject

\appendix


\section{The Coupling Kernel: Spectral Decomposition}
\label{app:coupling}

In this section we demonstrate how to identify the rank, $D$, of the coupling kernel $J(\phi,\phi')$ and show that this leads to  $\mathcal{O} (D \cdot N)$ couplings. The rank of the cosine coupling is two which is arguably the minimal rank coupling one can use for navigation. We also describe a strategy for engineering low-rank kernels with desirable characteristics and describe a method for approximating a general (full-rank) kernel with a lower rank kernel of dimension $D$.

We begin by considering the Hamiltonian in the absence of noise and embedded in a continuous context. Consider a measure $q(\phi)$ for active neurons; this can be explicitly related to the discrete target Hamiltonian by letting 
$$q(\phi) =\qv (\phi) \equiv  \sum_{k=1}^K n_k \delta(\phi-\thetav_k) .$$
In general we will think of $q(\phi)$ as a density (non-negative distribution in the space of measures) defined for $\phi \in [-\pi, \pi]$.
The noise-free Hamiltonian can now be written as a quadratic functional
$$H(q) = - \int_{\phi=-\pi}^\pi \int_{\phi'=-\pi}^\pi  q(\phi) J (\phi, \phi') q(\phi') \, d \phi'  \, d \phi $$
where $J$ is a symmetric kernel on a bounded domain. The reader can check that $H(\qv)$ yields the discrete Hamiltonian defined in \eqref{eq:H}. There are some strong parallels between this Hamiltonian and the analogous free energy and dynamical formulations for distributions of Kuramoto oscillators \cite{SakKur1986} and while the dynamics of the two systems are quite distinct, this previous work provides some insight into the dimensional reduction described below.

An important observation is that if we are looking to reduce the biological complexity of the system, the architecture of the animal's neural net (here represented by $J$) is fixed while the visual inputs will vary. As such, a reduction of the rank in $J$ is desirable if we wish to propose a simpler architecture. As $J$ is symmetric it is natural to consider a spectral decomposition for it; if the kernel is positive definite this is guaranteed to exist by Mercer's Theorem. More generally as the operator is on a compact domain and the kernel is bounded (from above and below), the spectrum is countable (possibly finite) and the spectral decomposition converges in an $L^2$ framework which we make precise below \cite{ReedSimon1981, polyanin2008handbook} 

Define the Hilbert-Schmidt integral operator,
\begin{equation}
\mathcal{T}[q] \equiv  \int_{\phi'=-\pi}^\pi  J(\phi, \phi') q(\phi') \ d \phi' \ \ .
\label{eq:HilSch}
\end{equation}
and also the inner-product and $L^2$ norm
$$\langle p (\phi), q (\phi)  \rangle 
\equiv \int_{\phi=-\pi}^{\pi} p(\phi) q (\phi) \, d \phi \ , \qquad \| q(\phi) \|^2 \equiv\langle q (\phi), q(\phi) \rangle .$$
In this norm, the operator is self-adjoint, which is a direct consequence of the symmetry of the Hamiltonian. To see this note that 
$$\langle p , T[q] \rangle =  \langle T[p] , q \rangle = - H\ .$$
As such, this operator has real eigenfunctions/eigenvalues $ \{ q_m(\phi) ;\lambda_m \}$ that satisfy 
$$T[q_m] = \lambda_m q_m \ \ .$$
As the kernel is bounded above and below (and the observing that the domain is compact), the set of eigenvalues, which we will denote $\Lambda$, is countable (and finite in many of the cases we consider here).  Let $D$ be the cardinality of $\Lambda$; we say a reduction is finite and of rank $D$ when this cardinality is finite.

For simplicity we normalize the eigenfunctions so that 
$$ \| q_m \|^2 =1 \ . $$ 
This allows us to write the spectral decomposition
\begin{equation}
\label{eq:spec_decomp}
J(\phi,\phi') = \sum_{\lambda_m \in \Lambda} \lambda_m 
q_m(\phi) q_m(\phi') .
\end{equation}
We will assume the kernel $J$ has mirror and interchange symmetries, 
$$J(\phi,\phi')=J(-\phi,-\phi') \qquad J(\phi,\phi')=J(\phi',\phi) \ . $$
As such, the operator $T$ has mirror symmetric in $\phi$; that is 
$$T[q(\phi)] = f(\phi) \qquad  \iff  \qquad T[q(-\phi)] = f(-\phi)$$
which allows us to break the eigenfunctions into odd and even functions.  Our goal here is to consider an approximation of $J$ where it is replaced by a small set of modes, often the even and odd eigenfunctions associated with the largest magnitude eigenvalues in the spectral decomposition yielding a rank two kernel. For the cosine kernel this answer is exact. 

If we wish to estimate the contribution to the energy from a particular set of modes, Parseval's identity is useful,
$$\|J(\phi, \phi') \|^2 \equiv 
\int_{\phi=-\pi}^{\pi} \int_{\phi'=-\pi}^{\pi} |J(\phi,\phi)|^2 \, d \phi' \ d \phi =
\sum_{\lambda_m \in \Lambda} \lambda_m^2   \ ,$$
where the last statement assumes the spectrum is complete in $L^2$ (which is true for square integrable $J$) \cite{ReedSimon1981}. In essence, Parseval tells us the relative contribution  of each eigenfunction to the $L^2$ norm of $J$ scales as the square of the associated eigenvalue.

\subsection{The cosine  kernel is rank two}
\label{app:cosine_kernel}
The most commonly used coupling \cite{Sridhar2021,Oscar2023,Gorbonos2024} is the cosine kernel
$$J(\phi , \phi') =\cos (\phi-\phi')$$
for which the complete spectrum is well known.
The two positive eigenvalues and eigenfunctions associated with $T$ are
$$\left \{q^+;\lambda_+ \right \} = \left \{ \frac 1 {\sqrt{\pi}} \cos\phi ; \pi \right \} \ ,  \quad \left \{q^-;\lambda_-  \right \} 
= \left \{ \frac 1 {\sqrt{\pi}} \sin\phi ; \pi \right \} \ ,
$$
where $q^+$ is even and $q^-$ is odd. 

The remaining eigenvalues of the Hilbert-Schmidt operator \eqref{eq:HilSch} are all zero; the associated eigenfunctions are
$$ \left \{q_0;\lambda_0  \right \} = \left \{ \frac 1 {\sqrt{2\pi}} ; 0 \right \} , $$ 
$$\left \{q^+_m;\lambda^+_m  \right \} =
 \left \{ \frac 1 {\sqrt{\pi}} \cos m\phi ; 0 \right \} \ , 
 \left \{q^-_m;\lambda^-_m  \right \} =
 \left \{ \frac 1 {\sqrt{\pi}} \sin m\phi ; 0 \right \} \ ,
\qquad m = 2,3,4, \cdots .$$
As these functions are in the null space of the operator  they don't contribute to the spectral decomposition which can be written as
\begin{align*}
  J(\phi , \phi') &=  \sum_{n=1}^{\infty} \lambda_m 
q_m(\phi) q_m(\phi') = \lambda^+q^+(\phi)q^+(\phi') + 
   \lambda^-q^-(\phi)q^-(\phi')  \ ,\\
&= 
\pi \left ( \frac 1 {\sqrt{\pi}} \cos\phi\right ) \left ( \frac 1 {\sqrt{\pi}} \cos\phi'\right ) 
+ \pi \left ( \frac 1 {\sqrt{\pi}} \sin\phi\right ) \left ( \frac 1 {\sqrt{\pi}} \sin\phi'\right ) 
\\
& =  \cos\phi   \cos\phi' +  \sin\phi \sin\phi'  \\
&=\cos (\phi-\phi') .
\end{align*}
This establishes that the cosine kernel is of rank 2.  Note that as $\lambda^{\pm}>0$ we can simplify the system a bit and prevent the proliferation of constants. Define
$$Q^+(\phi) = \sqrt{\lambda^+q^+(\phi)} \ , \quad
Q^-(\phi) = \sqrt{\lambda^-q^-(\phi)}
$$
which for the Kuramoto system $Q^+(\phi)=\cos(\phi)$ and $Q^-(\phi) = \sin \phi$. Then 
$$ J(\phi , \phi') = Q^+(\phi)Q^+(\phi') +Q^-(\phi)Q^-(\phi') \ . $$
Here the excitation of the even mode $Q^+$ measures \emph{coherence} while the odd mode $Q^-$ measures the \emph{torque} associated with a target configuration.

\subsection{An infinite dimensional example and its finite dimensional reduction}
\label{app:distort}
Unless a coupling kernel has some known structure or decomposition we expect it to have a countably infinite spectrum. Here we suggest a method for approximating it by a finite dimensional reduction. 

A previously used coupling kernel that generalizes the cosine kernel \cite{Sridhar2021,Oscar2023,Gorbonos2024} takes the form
$$J(\theta , \theta') = \mathcal{I}(\theta-\theta') $$
where
$$
\mathcal{I}(\phi) = 
\cos\left [\pi \left ( \frac{d_\mathbb{T}(\phi)} {\pi} \right )^\nu \right ]  \qquad d_\mathbb{T}(\phi) = \min_{m \in \mathbb{Z}} |\phi +2\pi m| \, 
$$
and which we will refer to as the \emph{angular distortion} coupling.
Here, the neural coupling, $J(\theta,\theta')$, depends only on the angular difference, $d_\mathbb{T}(\theta-\theta')$
of $\theta$ and $\theta'$ which is the smallest angle subtended by an arc between the points $\theta$ and $\theta'$ on the unit circle. For $\nu=1$ this reduces to the cosine coupling and for $\nu <1$ stretches the perceived angular difference between two targets. Following the previous authors, we note this when $\nu \ne 1$ the scaling is \emph{non-Euclidean} - that is angular differences are distorted. They also refer to the case where $\nu <1$ ($\nu>1$) as \emph{elliptic} (\emph{hyperbolic}) as proximal angles are stretched (shrunk).  For simplicity, we will examine this for $\nu=\frac 12$, which is a typical value used in these previous studies.

Neural couplings of the form $J(\theta , \theta') = \mathcal{I}(\theta-\theta')$ have multiple symmetries; they are symmetric, $2\pi$-periodic and invariant under rotation. 
Due to these symmetries the eigenfunctions of the Hilbert-Schmidt operator (\ref{eq:HilSch}) are just Fourier modes \cite{polyanin2008handbook}, 
$$ q_0  = \frac 1 {\sqrt{2\pi}} \ , \quad q^+_m  =
 \frac 1 {\sqrt{\pi}} \cos (m\phi)  \ , \quad 
 q^-_m = \frac 1 {\sqrt{\pi}} \sin (m\phi) \ ,
\qquad m = 1,2,3,\cdots .$$
The associated eigenvalues are
$$\lambda_m = \int_{-\pi}^{\pi} \mathcal{I}(\phi) \cos(m \phi) \, d \phi  $$ 
where $\lambda_0$ is associated to the constant mode $q_0$ and $\lambda_m$ for $m=1,2,3, \cdots$ is associated with the odd/even pair of eigenfunctions $q_m^+$ and $q_m^-$.

\begin{figure}[t!]
    \centering
    \includegraphics[width=\linewidth]{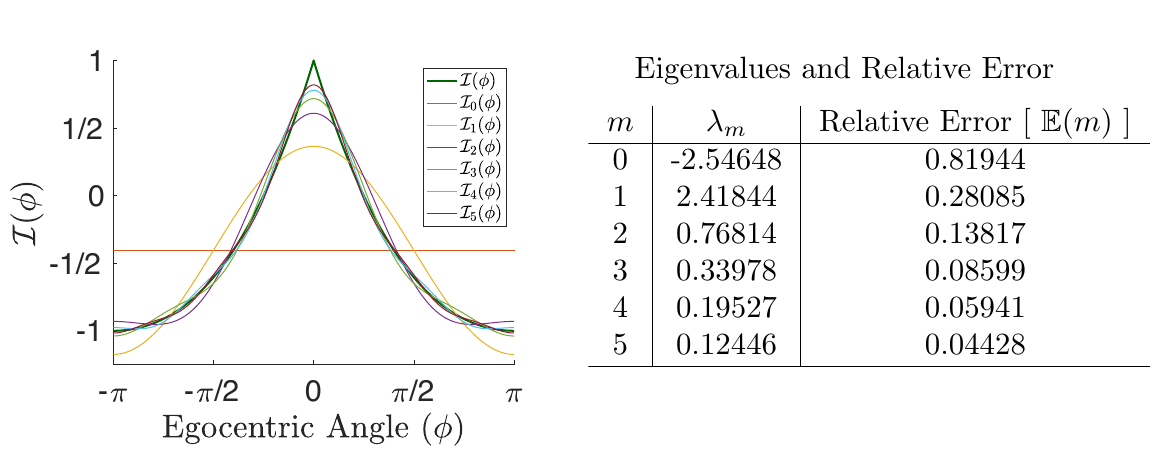}
    \caption{Convergence of the spectral approximation [$\mathcal{I}_m(\phi)$] to the distorted angle kernel. The figure illustrates how the $2m+1$ mode approximation converges to the kernel; pointwise the error is largest at the cusp.}
    \label{fig:spectrum_plot}
\end{figure}

We construct a $2M+1$ dimensional approximation to the neural coupling kernel
$$J_M(\phi,\phi') \equiv \lambda_0 q_0(\phi)q_0(\phi') + \sum_{m=1}^M \lambda_m \left [q^+_m(\phi)q^+_m(\phi') + q^-_m(\phi)q^-_m(\phi') \right ] $$
and define its relative $L^2$-error
\begin{equation}
\mathbb{E}(M) \equiv 
\frac{\|J(\phi,\phi)-J_m(\phi,\phi')\|}{\|J(\phi,\phi)\|}
= \frac{
\sqrt{\|J(\phi,\phi)\|^2-\lambda_0^2 -2\sum_{m=1}^M \lambda_m^2}
}
{\|J(\phi,\phi)\|}
\label{eq:RelErr}
\end{equation}
where the $L^2$ convergence of this spectral decomposition is equivalent to saying that $\lim_{M \to \infty} \mathbb{E}(M) =0$.

In Figure \ref{fig:spectrum_plot} we graph the approximations to $\mathcal{I}(\phi)$ defined as
$$\mathcal{I}_M(\phi)  \equiv \lambda_0 q_0(\phi) + \sum_{m=1}^M \lambda_m \left [q^+_m(\phi) + q^-_m(\phi) \right ] $$
while the qualitative convergence is good, the pointwise convergence at the peak is slow (due to the cusp). The adjacent table shows that the relative error for modes up to order 5 [$\mathbb{E}(5)$] is about 4.4 \% - the most important observation here is that the qualitative shape of the distorted angle kernel can be recovered by using a kernel with a few modes reducing the complexity of the coupling needed.

\section{Glauber Dynamics for Finite Dimensional Kernels}
\label{app:spectralGlauber}
For a coupling kernel of dimension $D$ the Glauber dynamics can be reduced to a $D$ dimensional system of ODEs. This reduction is known for the cosine kernel \cite{Sridhar2021,Gorbonos2024} but our method here applies to any finite dimensional kernel.
For simplicity, assume the eigenvalues $\Lambda = \{\lambda_1, \lambda_2, \cdots \lambda_D \}$ are positive which yields a spectral decomposition of the form
$$ J(\phi , \phi') = \sum_{j=1}^DQ_j(\phi)Q_j(\phi') \ \ .$$
Now define the \emph{spectral amplitudes} $\vec{\gamma} = \langle \gamma_1, \gamma_2, \cdots, \gamma_D \rangle$ where 
\begin{equation}
\gamma_j(\tau) = \langle \qv(\phi), Q_j(\phi) \rangle =\sum_{k=1}^K n_k (\tau) Q_j(\thetav_k)  \ .
\label{eq:gammaj}
\end{equation}
The Hamiltonian can now be written in terms of $\gamma_j$,
\begin{align*}
H &= - \int_{\phi=-\pi}^\pi \int_{\phi'=-\pi}^\pi  \qv(\phi) J (\phi, \phi') \qv(\phi') \, d \phi'  \, d \phi \\
&= 
- \int_{\phi=-\pi}^\pi \int_{\phi'=-\pi}^\pi  \qv(\phi) \left [ \sum_{j=1}^DQ_j(\phi)Q_j(\phi') \right ] \qv(\phi') \, d \phi'  \, d \phi\\
&= - \sum_{j=1}^D  \langle \qv(\phi), Q_j(\phi) \rangle  \langle \qv(\phi'), Q_j(\phi') \rangle \\
& = -\sum_{j=1}^D [\gamma_j(\tau)]^2 
\end{align*}

Differentiating \eqref{eq:gammaj} with respect to $\tau$ yields
\[
\frac{d \gamma_j}{d \tau} = \sum_{k=1}^K \frac{dn_k}{d\tau} Q_j(\thetav_k) + n_k\frac{d}{d\tau}Q_j(\thetav_k)\ .
\]
Assuming the observer and/or targets are not completely frozen in configuration space (e.g. see Section \ref{sec:navigation_model} for a navigation model based on the neural consensus), each $\thetav_k$ will be a function of time $t$ as each target changes its apparent angular location around the observer. However, if we assume a separation of timescales $\tau_0>>1$ between the fast neural dynamics $\tau=t/\tau_0$ and the external physical dynamics $t$, we can take $\frac{d}{d\tau}Q_j(\thetav_k)\rightarrow 0$. In this case, substituting \eqref{eq:glauberODE} yields
\begin{subequations}
\label{eq:gammaODE_ab}
\begin{gather}
\frac{d \gamma_j}{d \tau} =\mathcal{F}_j (\vec{\gamma}) \ , \qquad   
\mathcal{F}_j (\vec{\gamma}) \equiv
\sum_{k=1}^K \frac{\rho_k Q_j(\thetav_k)}
{1+\exp{\left( \beta (\Delta E)_k \right)} }- \gamma_j \qquad \textrm{for} \  j=1, 2, \cdots ,D
\label{eq:gammaODE_a} \\
\intertext{where}
(\Delta E)_k = \frac{\partial H}{\partial n_k} = - 2 \sum_{j=1}^D\gamma_j \frac{\partial \gamma_j}{\partial n_k } = - 2 \sum_{j=1}^D \gamma_j Q_j(\thetav_k) 
\label{eq:gammaODE_b} 
\end{gather}
\end{subequations}
which closes the system of ODEs for $\vec{\gamma}$. In Appendix \ref{app:gamstability} we solve for equilibrium solutions, $\vec{\gamma} = \vec{\gamma}^*$, to this system and examine their stability.

A curious observation is that the Glauber dynamics for $K$ targets is governed by the $K$ dimensional system \eqref{eq:glauberODE} while the system \eqref{eq:gammaODE_ab} above for $\vec{\gamma}$ is of dimension $D$. The values for the excitation weights $\vec{n}$ specify the spectral amplitudes $\vec{\gamma(t)}$ and when $K \le D$, the trajectories for $\vec{n}$ specify an invariant subspace of the trajectories for $\vec{\gamma(t)}$. However, when $D < K$ the evolution of $\vec{n}$ is under-constrained by the evolution of $\vec{\gamma}$; the crucial observation here is that the Hamiltonian $H$ and the changes in energy $(\Delta E)_k$ are specified by $\vec{\gamma}$ so $\frac{d n_k}{dt}$ is determined by the values of $n_k$ and $\vec{\gamma}$. One can think of the equations $\eqref{eq:glauberODE}$ as auxiliary equations that specified the detailed evolution of the excitation weights (which are constrained but under-determined by the spectral amplitudes $\vec{\gamma}$).

\section{Stability of Equilibrium Solutions}
\label{app:stability}
In this appendix, we collect the various stability calculations for both the fast neural dynamics and the navigational dynamics. We also call attention to where many of these results were derived for specific cases in the literature \cite{Sridhar2021,Gorbonos2024}.

\subsection{Glauber dynamics and stability for neural densities}
\label{app:ndstability}
In Section \ref{subsec:Glauber} we used Glauber dynamics to derive a set of $K$ ODEs for the  neural densities $\vec{n} = \langle n_1,n_2, \cdots n_K \rangle$ 
associated with each of the $K$ targets. The trajectories lie in the rectangular parallelepiped, $\mathcal{P}$, defined by $0 \le n_k \le \rho_k$.

\begin{equation}
\frac{dn_k}{d\tau} = F_k(\vec{n}) \ , \quad F_k(\vec{n}) = \frac{\rho_k}
{1+\exp{\left( \beta (\Delta E)_k \right)} }
- n_k  \ , \quad (\Delta E)_k = \frac{\partial H}{\partial n_k}  \qquad \textrm{for} \ k=1,2, \cdots K \ .    
\label{eq:app_glauberODE}
\end{equation}
Equilibrium solutions, $\vec{n} = \vec{n}^*$, satisfy the implicit equations 
\begin{equation}
	n_k^* = \frac{\rho_k}
{1+\exp{\left( \beta (\Delta E)_k^* \right)} } \ , \qquad (\Delta E)_k^* =
\left . \frac{\partial H}{\partial n_k} \right |_{\vec{n}=\vec{n}^*}
= - 2  \sum_{l=1}^K  n_\ell^* J(\thetav_k,\thetav_\ell).
\label{eq:app_steadystate}
\end{equation}
These equilibrium equations must be in the interior of $\mathcal{P}$ (as $\exp(\beta (\Delta E)_k^*) >0$).

\subsubsection{Lyapunov Stability}
We will use a Lyapunov function to show that generically this system will evolve to some minimizer satisfying these equilibrium conditions for $n_*$.
Statistical mechanics guides us to define the free energy
for this system as the internal energy minus the Gibb's entropy
$$G(\vec{n}) = H(\vec{n}) + \frac{1}{\beta} \sum_{j=1}^k \left[ n_j \ln n_j + (\rho_j - n_j) \ln(\rho_j - n_j) \right]$$ 
where $H$ and $G$ depend on $\vec{n}$.  Consider a trajectory with some initial condition in $\mathcal{P}$ evolving according to \eqref{eq:app_glauberODE}. Along this trajectory 
$$\frac{dG}{d \tau} = \sum_{j=1}^K \frac{\partial G}{\partial n_j} \frac{dn_j}{d \tau} = \sum_{j=1}^K \frac{\partial G}{\partial n_j} F_j(\vec{n})
$$
where
\begin{align*}
\frac{\partial G}{\partial n_j} &= \frac{\partial H}{\partial n_j} + \frac{1}{\beta} \left( \ln n_j + 1 - \ln(\rho_j - n_j) - 1 \right)\\
&= (\Delta E)_j + \frac{1}{\beta} 
\ln \left(\frac{n_j}{\rho_j - n_j}\right ) \\
&= \frac{1}{\beta} \left [ \beta(\Delta E)_j 
- \ln\left(\frac{\rho_j - n_j}{n_j} \right )\right ] \\
&=\frac{1}{\beta} \left [ \mathcal{A}_j -\mathcal{B}_j \right ]
\end{align*}
with $ \mathcal{A}_j =\beta(\Delta E)_j   $ and $\mathcal{B}_j =  \ln\left(\frac{\rho_j - n_j}{n_j} \right )$ . Also 
\begin{align*}
F_j(\vec{n}) &= \frac{\rho_j}
{1+\exp{\left( \beta (\Delta E)_j \right)} }
- n_j  \ , \\
 &= \frac{\rho_j-n_j -n_j \exp{\left( \beta (\Delta E)_j \right ) } }
{1+\exp{\left( \beta (\Delta E)_j \right)} }\\
&=\frac{n_j} {1+\exp{\left( \beta (\Delta E)_j \right)} }
\cdot \left [ \frac{\rho_j-n_j}{n_j} - \exp{\left( \beta (\Delta E)_j \right)}       \right ] \ , \\
&=- \frac{n_j} {1+\exp{\left( \beta (\Delta E)_j \right)} } \cdot \left [e^{\mathcal{A}_j} -e^{\mathcal{B}_j} \right ] \ .
\end{align*}
Multiplying them together yields 
$$
\frac{dG}{d \tau} = - \sum_{j=1}^{K} \frac{n_j/\beta} {1+\exp{\left( \beta (\Delta E)_j \right)} } \cdot \left [ \mathcal{A}_j -\mathcal{B}_j \right ]
\cdot \left [e^{\mathcal{A}_j} -e^{\mathcal{B}_j} \right ]
$$
As the exponential is a monotonically increasing function, we find $\frac{dG}{dt} \le 0$ with equality if and only if $\mathcal{A}_j=\mathcal{B}_j$ for $j=1,2, \cdots K$. This condition is equivalent to the implicit steady-state equations \eqref{eq:app_steadystate} as previously claimed.

\subsubsection{Linear stability of neural density equilibria}
We can use the Lyapunov function to determine the linear stability of the equilibria $\vec{n}^*$ and show that the associated spectrum is real. The Hessian of $G$ (evaluated at $\vec{n}=\vec{n}^*$)  is given by
\begin{equation}
\label{eq:nHessian}
   {Q}_{\ell k} =   \left .  \frac{\partial^2 G}{\partial n_\ell \partial n_k} \right |_{\vec{n}=\vec{n}^*} =  -2 J(\thetav_\ell,\thetav_k) 
+ \frac {\delta_{\ell k}}{\beta} \left [ \frac{1}{n_\ell^*} + \frac{1}{\rho_\ell- n_\ell^*}  \right ] 
\end{equation} 
where $\delta_{\ell k}$ is the Kronecker $\delta$-function. The Hessian by definition is symmetric (implying real eigenvalues) and the condition for stability is that all its eigenvalues are positive. We can show this also specifies the spectrum for the linear stability problem associated with the ODE system. Let
$$ n_j( \tau) = n^*_j +\varepsilon \eta_j(\tau)  \ ,$$
and linearizing in $\varepsilon$ yields 
$$\frac{d\eta_j}{d \tau} = W_{jk} \, \eta_k$$
where
$$W_{jk} = \left . \frac{\partial F_j}{\partial n_k} 
\right |_{\vec{n}=\vec{n}^*} \ .$$
Define a diagonal matrix
$$\mathcal{D}_{\ell j} = \frac{1}{\beta}   \left [ \frac{1}{n_j^*} + \frac{1}{\rho_j- n_j^*}  \right ] \delta_{\ell j}$$
where we note that the entries are all positive for $\beta>0$.
Then a tedious calculation yields
$${Q}_{\ell k} = -\mathcal{D}_{\ell j}W_{jk}$$
at $\vec{n}=\vec{n}^*$ assuring us the spectrum of the linear stability problem is the negative of that of for the corresponding critical points of the Lyapunov functional and therefore the stability criteria are identical.

\subsection{Equilibria and stability for the \texorpdfstring{$\vec{\gamma}$}{gamma} system}
\label{app:gamstability}
In Section \ref{app:spectralGlauber} we derived a set of $D$ ODEs for the spectral amplitudes
$\vec{\gamma} = \langle \gamma_1, \gamma_2, \cdots \gamma_D \rangle$,
\begin{equation}
\label{eq:app_gammaODE}
\frac{d \gamma_j}{d \tau} =\mathcal{F}_j (\vec{\gamma}) \ , \qquad   
\mathcal{F}_j (\vec{\gamma}) \equiv
\sum_{k=1}^K \frac{\rho_k Q_j(\thetav_k)}
{1+\exp{\left( \beta (\Delta E)_k \right)} }- \gamma_j , \qquad 
(\Delta E)_k = - 2 \sum_{j=1}^D \gamma_j Q_j(\thetav_k) 
\end{equation}
for $j=1, 2, \cdots ,D$.
The equilibrium solutions, $\vec{\gamma} = \vec{\gamma}^*$, can be solved for in a manner akin to the solution to \ref{eq:app_glauberODE} yielding an implicit solution
\begin{equation}
\label{eq:gamma_steadystate}
	\gamma_j^* = \sum_{k=1}^K \frac{\rho_k Q_j(\thetav_k)}
{1+\exp{\left( \beta (\Delta E)_k^* \right)} }
\qquad 
(\Delta E)_k^* = - 2 \sum_{j=1}^D \gamma_j^* Q_j(\thetav_k) \ \ .
\end{equation}
The system \eqref{eq:gammaODE} is a gradient flow (which was essentially derived as a free energy by \cite{Gorbonos2024}) which facilitates the stability calculation of the equilibria. Let
$$\mathcal{G} = \frac 1 2 \sum_{j=1}^D (\gamma_j)^2 -
\frac {1} { 2 \beta} \sum_{k=1}^K \rho_k\ln 
\left [ 1 + \exp (-\beta ( \Delta E)_k ) \right ] $$
then
$$ \frac{d \gamma_j}{d \tau} = - \frac{\partial \mathcal{G} } {\partial \gamma_j} .$$
Stability of an equilibrium solution $\vec{\gamma}^*$ is governed by Hessian of $G$,
\begin{align*}
\mathcal{Q}_{j \ell} &=   \left .  \frac{\partial^2 \mathcal{G}}{\partial \gamma_j \partial \gamma_\ell} \right |_{\vec{\gamma}=\vec{\gamma}^*} \\
&= 
\delta_{j \ell}
- 2 \beta \sum_{k=1}^K \rho_k   Q_j(\thetav_k) Q_\ell(\thetav_k)
\frac{\exp [\beta ( \Delta E)^*_k]}
{\left [ 1 + \exp (\beta ( \Delta E)^*_k  \right ]^2} \\
&=
\delta_{j \ell}
-  \frac{\beta}{2} \sum_{k=1}^K \rho_k   Q_j(\thetav_k) Q_\ell(\thetav_k) \,
\textrm{sech}^2 \left  (\frac{\beta ( \Delta E)^*_k}{2} \right )
\end{align*}
As with \eqref{eq:nHessian}, the stability of $\vec{\gamma}^*$ is determined by the eigenvalues of the Hessian whose symmetry guarantees a real spectrum. Here negative eigenvalues correspond to instability.

\subsection{\texorpdfstring{$\gamma$}{gamma}-stability for the cosine kernel}
\label{sec:gamma_stability_cosine}
For the cosine neural coupling, we have 
$Q_+(\phi)= \cos(\phi)$ , $Q_-(\phi) = \sin(\phi)$, and from \eqref{eq:Gammastar} we see that 
$$(\Delta E)_{k}^* = - 2 R^* \cos(\thetav_k -\Theta^*) \ .$$
The Hessian becomes a $2 \times 2$ matrix 
\begin{equation}
    \label{app:Hess}
\mathcal{Q} =
\begin{bmatrix}
1-  \displaystyle \frac{\beta}{2} \sum_{k=1}^K \rho_k   \cos^2(\thetav_k)  \,
\textrm{sech}^2 \left  (\frac{\beta ( \Delta E)^*_k}{2} \right ) 
&-  \displaystyle \frac{\beta}{2} \sum_{k=1}^K \rho_k   \cos(\thetav_k) \sin(\thetav_k) \,
\textrm{sech}^2 \left  (\frac{\beta ( \Delta E)^*_k}{2} \right )  \\
-  \displaystyle \frac{\beta}{2} \sum_{k=1}^K \rho_k   \cos(\thetav_k) \sin(\thetav_k) \,
\textrm{sech}^2 \left  (\frac{\beta ( \Delta E)^*_k}{2} \right )  
&1-  \displaystyle \frac{\beta}{2} \sum_{k=1}^K \rho_k   \sin^2(\thetav_k)  \,
\textrm{sech}^2 \left  (\frac{\beta ( \Delta E)^*_k}{2} \right ) 
\end{bmatrix}
\end{equation}
A necessary and sufficient condition for stability (that is for $\mathcal{Q}$ to be positive definite) is that the diagonal elements be positive ($\mathcal{Q}_{11}>0$,$\mathcal{Q}_{22}>0$), that is 
\begin{subequations}
\label{eq:gamma_stab_diag}
\begin{align}
1 & >  \displaystyle \frac{\beta}{2} \sum_{k=1}^K \rho_k   \cos^2(\thetav_k)  \,
\textrm{sech}^2 \left  (\frac{\beta ( \Delta E)^*_k}{2} \right ) \ , 
\label{eq:gamma_stab_diagA}
\\
1 & > \displaystyle \frac{\beta}{2} \sum_{k=1}^K \rho_k   \sin^2(\thetav_k)  \,
\textrm{sech}^2 \left  (\frac{\beta ( \Delta E)^*_k}{2} \right ) \ .
\label{eq:gamma_stab_diagB}
\end{align}
\end{subequations}
and that $\det{\mathcal{Q}} >0$.
In the case where the configuration has mirror symmetry,
$\mathcal{Q}_{12} =\mathcal{Q}_{21} =0 $
and conditions \eqref{eq:gamma_stab_diag} are necessary and sufficient.

For a $\gamma$-equilibrium solution at $\Theta*=0$ (that is the neural consensus aligns with the observer's bearing) $\gamma^*$ is real. In this case, the second condition \eqref{eq:gamma_stab_diagB} corresponds to stability to perturbations that purely angular (that is perpendicular to the direction of motion) and  was derived by Gorbonos (\cite{Gorbonos2024}, Eqn. 10) for targets of equal weight.


\subsection{Navigational stability}
\label{app:navstability}
Here we introduce the concept of \emph{navigational stability}.
Consider a  $\gamma$-equilibrium where the neural consensus angle aligns with the walker's bearing; that is where $\gamma^*$ is real or equivalently where $\Theta^*=0$. We can ask if this equilibrium is stable in a navigational sense; that is will small perturbations in the walker's orientation and neural state decay?

To answer this question we need to examine the evolution of the walker's orientation in the allocentric frame.
For the deterministic torque model \eqref{eq:walkerPhi} written in the allocentric frame
\begin{equation}
\frac{d \bearing}{dt} =  \Krate\Rbar \sin(\thetabar/2),\label{app:walkerPhi}
\end{equation}
we say that when an walker has a \emph{navigational equilibrium} at an allocentric angle $\bearing(t) =\bearing^*$ when the neural consensus angle has $\Theta^*=0$ (that is where the torque is zero).

To examine the stability of this navigational equilibrium we first assume that $\gamma^*$ is a $\gamma$-stable equilibrium. Now, consider \eqref{app:walkerPhi} in the allocentric frame. We write $\bearing(t) =\bearing^* +\epsilon \xi(t)$ and note that $R^*$ and $\Theta^*$ depend on $\bearing$ as the location and the weighting of the targets depend upon the orientation of the egocentric reference frame. Linearizing in $\epsilon$ yields
\begin{align*}
    \Theta^*(\bearing^* +\epsilon \xi) &=   \Theta^*(\bearing^*) + \epsilon \xi
    \left . \frac{\partial \Theta^*}{\partial \bearing^{\ }} \right |_{\bearing =\bearing^*} + \bigoh(\epsilon^2) 
    = \epsilon C\xi + \bigoh(\epsilon^2)  \qquad C = \left . \frac{\partial \Theta^*}{\partial \bearing^{\ }} \right |_{\bearing =\bearing^*} \\
 \Rbar(\bearing^* +\epsilon \xi) &=   \Rbar(\bearing^*) + \bigoh(\epsilon) 
\end{align*}
Linearizing \eqref{app:walkerPhi} yields at $\bigoh(\epsilon)$
$$ \frac{d \xi}{dt} =   \frac{C \Krate\Rbar}{2} \xi \ ,\label{app:lintorquemodel}$$
so the stability depends solely on the sign of $C$.

When the weighting of targets is uniform and space is not warped then $\thetabar$ corresponds to a fixed location in the allocentric frame. The allocentric angles can now be related to the egocentric angles by noting that
 \begin{equation}
     \theta = \varphi-\bearing(t)  \ . 
 \end{equation}
Then as $\thetabar$ is in the allocentric frame, 
$$\thetabar(\bearing^*+ \epsilon \xi ) = \thetabar(\bearing^*) - \epsilon \xi \qquad \Rightarrow \qquad C=-1 $$
which tells us that if a navigational equilibrium is $\gamma$-stable it is navigationally stable. This applies to both the unwarped, unweighted cosine coupling and the distorted angle model of Gorbonos \cite{Gorbonos2024}.

When space is weighted and/or warped this calculation becomes significantly more difficult. For simplicity we consider only the cosine kernel which may be potentially warped or weighted. In this case we know that the $\gamma$-equilibrium satisfies the implicit equation \eqref{eq:Gammastar} 
\begin{equation}
\label{app:Gammastar}
\gamma^* = \sum_{k=1}^K \frac{\rho_k e^{i\thetav_k}}
{1+\exp{\left( -2\beta \Real [\gamma^* e^{-i \thetav_k}] \right)} } ,
\end{equation}
where $\thetav_k$ and $\rho_k$ may depend upon the walker's bearing in the cases of warping and weighting respectively. To connect this to the linear $\gamma$-stability problem in \eqref{sec:gamma_stability_cosine} we write this in terms of real and imaginary parts of $\gamma^*$. Remembering that 
$$ R^*e^{i \Theta^*} =\gamma^* = \gamma^*_+ + i \gamma^*_- $$
and differentiating with respect to $\bearing$ yields
$$\left [ \frac{\partial R^*}{\partial \Phi} + i R^* \frac{\partial \Theta^*}{\partial \Phi} \right ] e^{i \Theta^*} =
\frac{\partial \gamma^*_+}{\partial \Phi} + i \frac{\partial \gamma^*_-}{\partial \Phi}  \ .$$
Evaluating at the navigational equilibrium where $\Theta^*=0$ and taking the imaginary part yields
$$C =\left . \frac{\partial \Theta^*}{\partial \bearing^{\ }} \right |_{\bearing =\bearing^*} = \left . \frac {1}{R^*} \frac{\partial \gamma^*_-}{\partial \Phi}\right |_{\bearing =\bearing^*}  \\ .$$
In addition we note that $\gamma^*_+ =R^*$ and $\gamma^*_-=0$ at the navigational equilibrium.

Taking the real and imaginary part of \eqref{app:Gammastar} yields
\begin{align}
0 &=\gamma^*_+ - \sum_{k=1}^K \frac{\rho_k \cos {\thetav_k}}
{1+\exp{\left ( -2\beta [\gamma^*_+ \cos(\thetav_k)+\gamma^*_- \sin(\thetav_k) ] \right)} } \ , \\
0 &=\gamma^*_- - \sum_{k=1}^K \frac{\rho_k \sin {\thetav_k}}
{1+\exp{\left ( -2\beta [\gamma^*_+ \cos(\thetav_k)+\gamma^*_- \sin(\thetav_k) ] \right)} } \ .
\label{app:gammastar_real_imag}
\end{align}
In the allocentric coordinate system, if we assume that the $k^{th}$ target is at $\varphi_k$ then 
$$\thetav_k = \mathcal{U}(\theta_k) =  \mathcal{U}(\varphi_k - \Phi) $$
and then we can define
\begin{equation}
\thetav_k' 
\equiv \left . \frac{\partial \thetav_k}{\partial \bearing^{\ }} \right |_{\bearing =\bearing^*} 
= \begin{cases}
-\mu(\varphi_k)  & \textrm{warped neural field} \\
-1 &  \textrm{unwarped neural field} \\
\end{cases}
\end{equation}
where the unwarped coordinate system is equivalent to choosing $\mathcal{U}(\varphi)=\varphi$ which yields
$\thetav_k' = -1 $. Similarly, we can define
\begin{equation}
\rho_k' 
\equiv \left . \frac{\partial \rho_k}{\partial \bearing^{\ }} \right |_{\bearing =\bearing^*} 
= \begin{cases}
-\mu'(\varphi_k)  & \textrm{weighted targets} \\
0 &  \textrm{unweighted targets} \\
\end{cases}
\end{equation}
where for targets weighted by neural density we remember that  
$\rho_k = \mu(\theta_k) =\mu(\varphi_k-\Phi)$.

Differentiating \eqref{app:gammastar_real_imag} with respect to $\bearing$ and evaluating at $\bearing=\bearing_*$ yields an inhomogeneous linear system.  Define
$$
 \gamma_+' = \left .  \frac{\partial \gamma^*_+}{\partial \Phi}\right |_{\bearing =\bearing^*}  \quad , \qquad 
 \gamma_-' = \left .  \frac{\partial \gamma^*_-}{\partial \Phi}\right |_{\bearing =\bearing^*} \quad , \qquad  
 \vec{\gamma}' =
\begin{bmatrix}
 \gamma_+' \\
\gamma_-'
\end{bmatrix} \ ,
$$
the coefficient matrix
$$\mathcal{M} =
\begin{bmatrix}
M_{11}& M_{12}\\
M_{21}& M_{22}
\end{bmatrix} =
\begin{bmatrix}
1-  \displaystyle \frac{\beta}{2} \sum_{k=1}^K \rho_k   \cos^2(\thetav_k)  \,
\textrm{sech}^2 \left[ \beta R^* \cos(\thetav_k ) \right ] 
&-  \displaystyle \frac{\beta}{2} \sum_{k=1}^K \rho_k   \cos(\thetav_k) \sin(\thetav_k) \,
\textrm{sech}^2 \left[ \beta R^* \cos(\thetav_k ) \right ]  \\
-  \displaystyle \frac{\beta}{2} \sum_{k=1}^K \rho_k   \cos(\thetav_k) \sin(\thetav_k) \,
\textrm{sech}^2 \left[ \beta R^* \cos(\thetav_k ) \right ]   
&1-  \displaystyle \frac{\beta}{2} \sum_{k=1}^K \rho_k   \sin^2(\thetav_k)  \,
\textrm{sech}^2 \left[ \beta R^* \cos(\thetav_k ) \right ] 
\end{bmatrix}
\ ,
$$
and the vector 
\begin{align*}
\vec{b} = 
\begin{bmatrix}
    b_1 \\ b_2 
\end{bmatrix}
= &  \sum_{k=1}^K 
\frac{\rho_k\theta_k'}
{1+\exp{\left [ -2\beta R \cos(\thetav_k) \right ]} }
\begin{bmatrix}
- \sin {\thetav_k}
\\
\phantom{-} \cos {\thetav_k}
\end{bmatrix}
-
\frac {\beta} 2 { \rho_k \theta_k' \sin(\thetav_k) }
{ \textrm{sech}^2 \left[ \beta R^* \cos(\thetav_k ) \right ] } 
\begin{bmatrix}
\cos {\thetav_k}
\\
\sin {\thetav_k}
\end{bmatrix} \\
&+  \sum_{k=1}^K 
 \frac{\rho_k' }
{1+\exp{\left [ -2\beta R \cos(\thetav_k) \right ]} }
\begin{bmatrix}
\cos {\thetav_k}
\\
\sin {\thetav_k}
\end{bmatrix} \ \ .
\end{align*}
Then $\vec{\gamma}'$ satisfies the linear equation
$$
\mathcal{M} \vec{\gamma}'
= \vec{b} \ .
$$
We can recover the unwarped, unweighted result stated above by letting $\thetav_k'=-1$  and $\rho_k'=0$. Then the system has a solution $\vec{\gamma}' = [0 , -R^* ]^T$ which yields $C=-1$  as previously advertised.

For the general system we first note that $\mathcal{M}$ is exactly the Hessian \eqref{app:Hess} and must be positive definite for $\gamma$-stability. Navigational stability requires $C = \gamma_-'/\Rbar  <0$. From Cramer's rule
$$ \gamma_-' = \frac 1 {\det{\mathcal{M}}}
{
\begin{vmatrix}
   {M}_{11} & b_1 \\
  {M}_{21} & b_2
\end{vmatrix}
} $$
but as $\mathcal{M}$ is positive definite, $\det{\mathcal{M}} >0$ so the conditions for navigational stability are that (i) The equilibrium is $\gamma$-stable and (ii) the condition
$$ {M}_{11} b_2 < {M}_{21} b_1 \ .  $$
which guarantee that $\gamma_-'$ and $C$ are negative.

\section{Additional Figures}
\label{app:model_details}

In this section, we include additional figures that are aimed at providing further examples of model dynamics and demonstrating how different model parameterizations can affect the geometry of decision making. 

\begin{figure}[t]
    \centering
    \includegraphics[width=\linewidth]{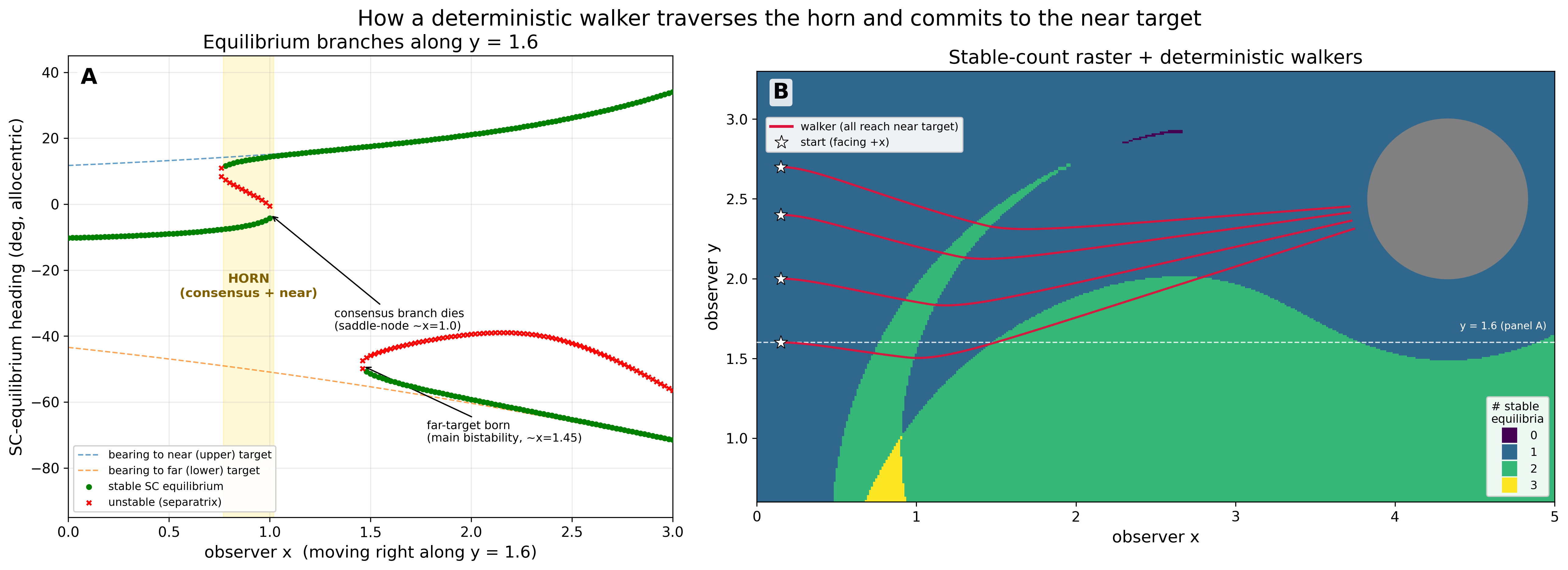}
    \caption{Bifurcation diagrams for off-axis deterministic walker dynamics with two circular targets, with specific focus on the ``horns.'' A linear cutoff function with $a=\pi/8$, $b=\pi$, $\beta=10$ was used, with no angle weight. Targets were the same as the fly geometry: radius 0.5 at locations $(4.33,\pm 2.5)$. (A) Self-consistent equilibrium directions (green: stable, red: unstable) along the $y=1.6$ slice of $(x,y)$-parameter space as $x$ changes. Dotted lines correspond to the angles at which the targets are located for a given value of $x$. Moving left to right, a walker will be drawn first to the single, consensus equilibrium with this behavior maintained through the bi-stable horn. On the other side of the horn, only the near-target equilibrium remains until, depending on the trajectory, a saddle-node bifurcation gives rise to an equilibrium corresponding to the other target. This corresponds to a potential passage through the main green region in (B). However, without a significant perturbation, this will not affect walker dynamics because the walker has already been drawn to the near-target equilibrium branch and is nowhere near the basin of attraction for the far-target equilibrium.}
    \label{fig:horn_dynamics}
\end{figure}

In Fig. \ref{fig:horn_dynamics}, we zoom into the dynamics of the horn-shaped regions of bifurcation space that are typical for targets of finite size. For an observer to the left of the horn arc, the single stable self consistent (SC) equilibrium points toward a consensus location between the two targets. A walker would then follow that direction into the green horn region, where a fold bifurcation gives rise to a second SC-equilibrium point indicating the near target. However, if the walker entered this region tracking the consensus direction with low angular noise, it will continue following the consensus equilibrium until it becomes unstable in a backward fold bifurcation at the other side of the horn, at which point it will switch toward the direction of the near target. Thus, the horn regions are regions of hysteresis in $(\Theta,R)$ space.

Walker trajectories that move across the horn can also transit the main bistable region as seen by the lower track in Fig. \ref{fig:horn_dynamics}B. The boundary of this main region is a saddle-node bifurcation away from the stable equilibrium corresponding to the near target; the new stable SC-equilibrium that is born corresponds to the far target. However, a walker will continue toward the near target unless something perturbs it heavily so that it falls into the new basin of attraction.

\begin{figure}
    \centering
    \includegraphics[width=\linewidth]{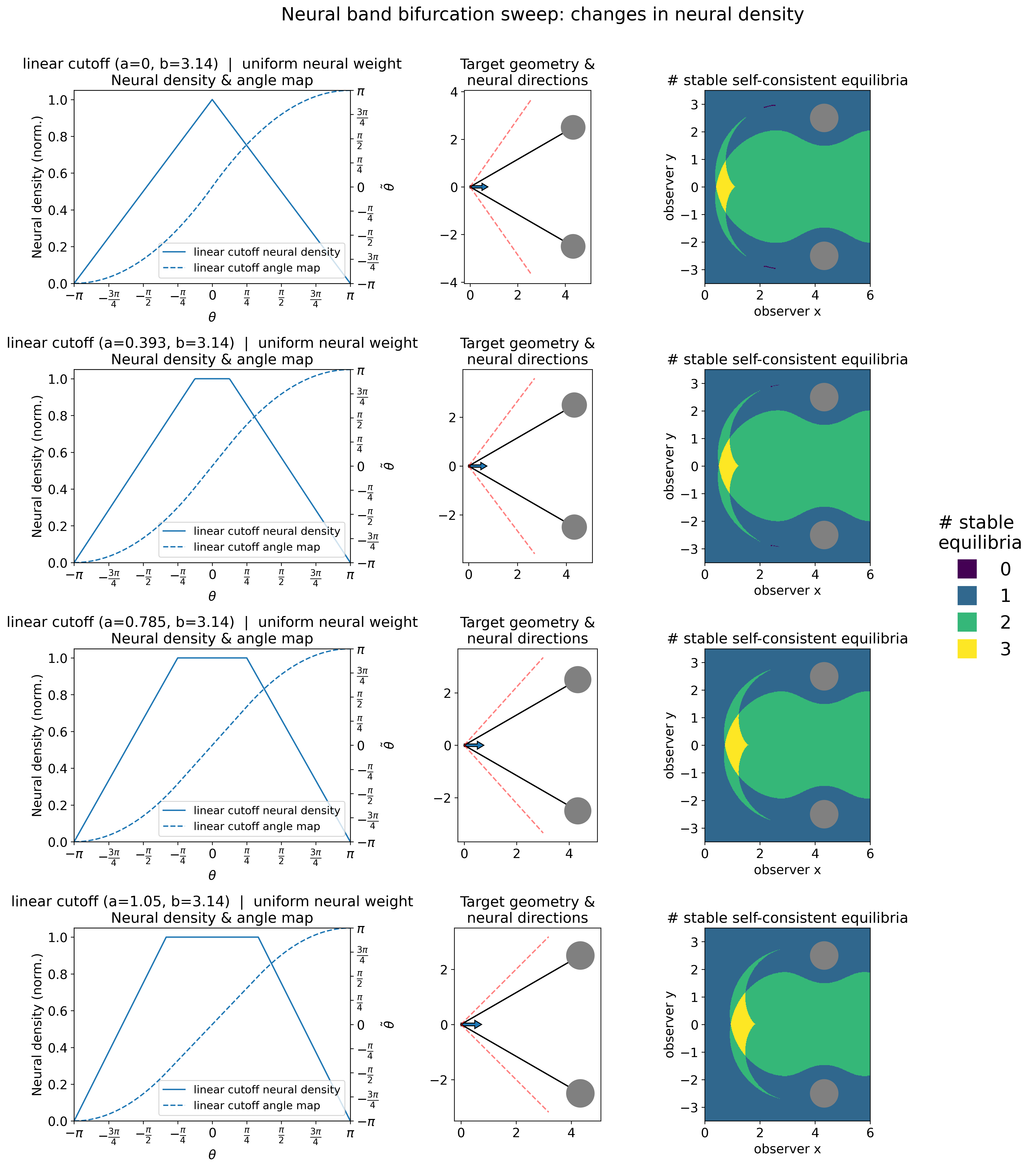}
    \caption{Neural direction geometry and bifurcation structure as parameter $a$ changes in the linear cutoff function $\mu(\theta)$ with no weighting of the targets by angular location (uniform weighting function $\omega(\theta)$). First column: visualization of the linear cutoff density function as $a$ varies and $b$ is held constant at $\pi$, together with the corresponding angle map (the integral of linear cutoff neural density function). Second column: for an observer located at the origin facing along the positive $x$-axis, the physical target geometry versus the corresponding neural angles of each target (red dashed lines). Third column: the number of stable self-consistent equilibria as a function of spatial location. Note that increasing the $a$ parameter brings the bifurcation structure closer to the targets. Also, the 0-stable islands at the ends of the horn structures are often thin and require a high level of resolution to capture; their absence in any given plot should not be taken to indicate their complete absence in reality.}
    \label{fig:sweep_a_uniform_weight}
\end{figure}

\begin{figure}
    \centering
    \includegraphics[width=\linewidth]{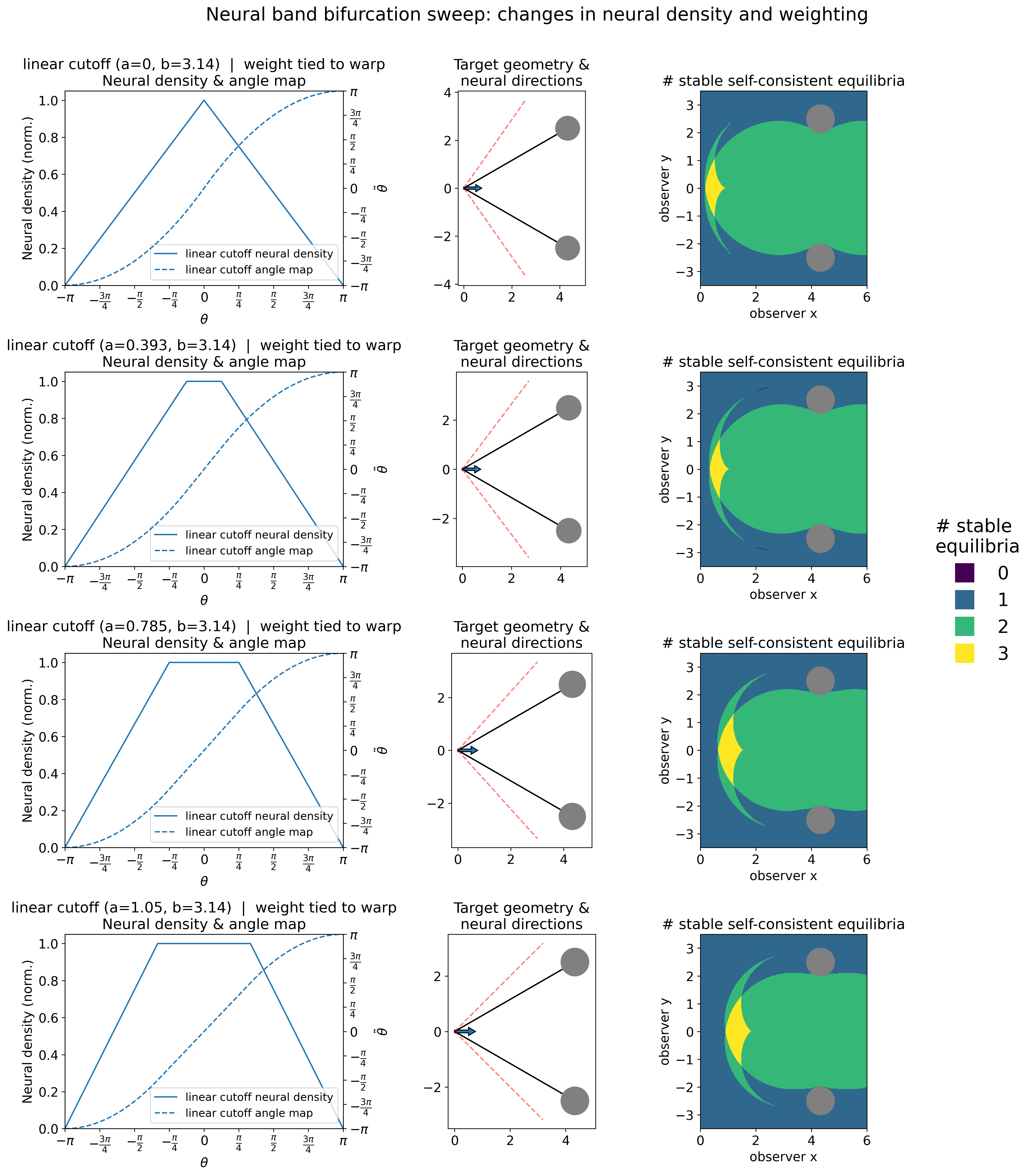}
    \caption{Neural direction geometry and bifurcation structure as parameter $a$ changes in the linear cutoff function $\mu(\theta)$ with weighting of the targets given by $\omega(\theta)=\mu(\theta)$. First column: visualization of the linear cutoff density function as $a$ varies and $b$ is held constant at $\pi$, together with the corresponding angle map (the integral of linear cutoff neural density function). Second column: for an observer located at the origin facing along the positive $x$-axis, the physical target geometry versus the corresponding neural angles of each target (red dashed lines). Third column: the number of stable self-consistent equilibria as a function of spatial location. Note that compared to Fig. \ref{fig:sweep_a_uniform_weight}, the bifurcation structure is pushed slightly back from the targets ($\mu(\theta)$ biases the observer toward targets in the center of the visual field, resulting in increased discrimination so that decisions are made earlier) and the interior bi-stable region has increased in size. As before, the 0-stable islands at the ends of the horn structures are often thin and require a high level of resolution to capture; their absence in any given plot should not be taken to indicate their complete absence in reality.}
    \label{fig:sweep_a_tied_weight}
\end{figure}

\begin{figure}
    \centering
    \includegraphics[width=\linewidth]{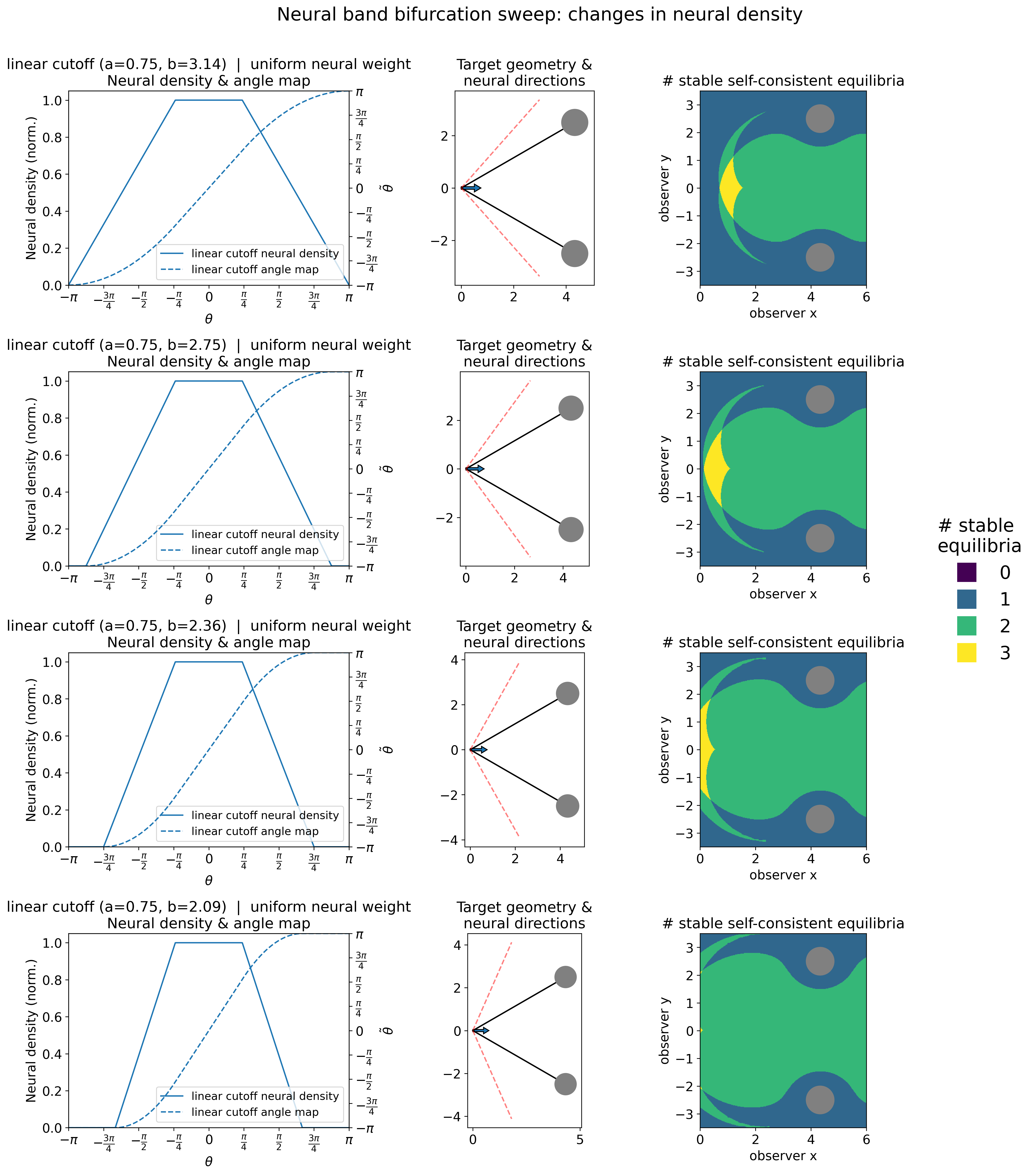}
    \caption{Neural direction geometry and bifurcation structure as parameter $b$ changes in the linear cutoff function $\mu(\theta)$ with no weighting of the targets by angular location (uniform weighting function $\omega(\theta)$). First column: visualization of the linear cutoff density function as $b$ varies and $a$ is held constant at 0.75, together with the corresponding angle map (the integral of linear cutoff neural density function). Second column: for an observer located at the origin facing along the positive $x$-axis, the physical target geometry versus the corresponding neural angles of each target (red dashed lines). Third column: the number of stable self-consistent equilibria as a function of spatial location. Note that decreasing the $b$ parameter (larger blind spots) pushes the bifurcation structure further out from the targets. Absence of the 0-stable islands at the ends of the horn structures in these plots may simply be due to how thin they are compared with the region as a whole and should not be taken as diagnostic.}
    \label{fig:sweep_b_uniform_weight}
\end{figure}

Figs. \ref{fig:sweep_a_uniform_weight}-\ref{fig:sweep_b_uniform_weight} demonstrate how changes in the neural density function and neural weighting affect the geometry neural direction and bifurcation space. Fig. \ref{fig:sweep_a_uniform_weight} adjusts the point where the neural density begins to fall off while keeping the field of view constant at $(-\pi,\pi)$. Scanning up the middle column, one can see that neural angles are pushed further out from the original target directions for more peaked distributions resulting in a regime where decisions are made further away from the target. In other words: more peaked neural distributions result in wider neural angles which correspond to more discrimination between targets. 

Fig. \ref{fig:sweep_a_uniform_weight} is the same plot but with neural group weighting tied to the neural density function as well. This means that observers experience a foveal effect: they will be more drawn to targets in front of them according to the neural density function. The main effect on bifurcation space is that the primary region of bistability now extends fully from one target to the next since an observer near one target but facing the other is now drawn to the far target it is facing rather than the near target behind them. A secondary effect is that the horns are generally a bit shorter.

Fig. \ref{fig:sweep_b_uniform_weight} returns to the uniform weighting of Fig. \ref{fig:sweep_a_uniform_weight} but varies the extent of the neural density function instead of its peakedness. This greatly affects how early a decision is made by a walker approaching the targets, suggesting that animals with narrower fields of vision are better equipped to discriminate against targets while further away from them compared to animals with fields of view closer to 360 degrees.

\begin{figure}
    \centering
    \includegraphics[width=0.6\linewidth]{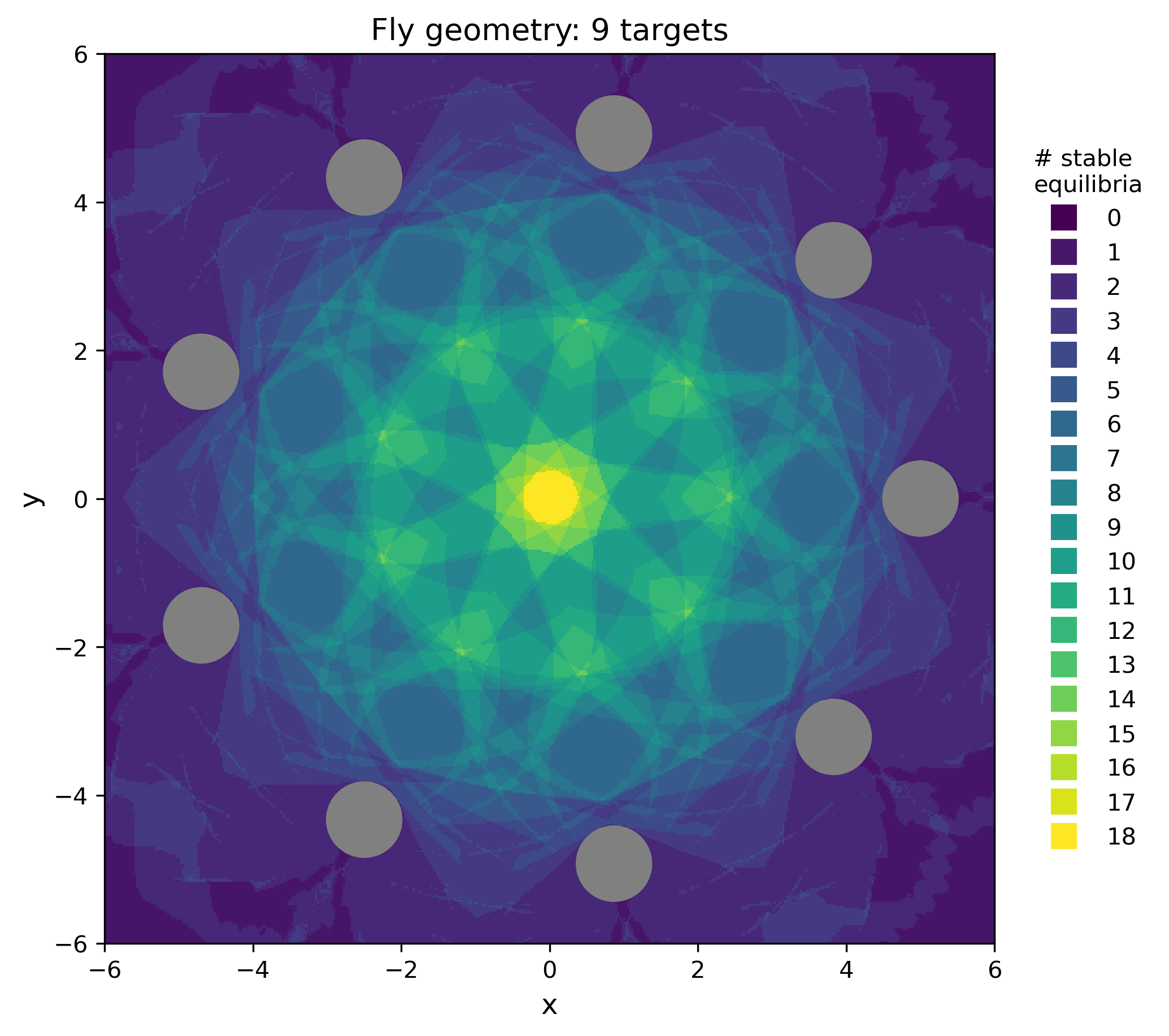}
    \caption{Bifurcation diagram expanded to 9 targets based on the target size and geometry used in \cite{Sridhar2021}. Model dynamics were not fit to the experimental data; instead, an identical linear cutoff function was used for both $\mu(\theta)$ and $\omega(\theta)$ with $a=0.25\pi$, $b=0.9\pi$, and $\beta=45$ in order to better demonstrate the bifurcation structure. With full symmetry and random starting angle, the bias toward any particular target disappears.}
    \label{fig:9target}
\end{figure}

Fig. \ref{fig:9target} demonstrates the rotational symmetry in decision-making space that results from having a complete circle of targets, in contrast to partial circle constructions (e.g. the 3-target case) where outer targets are disadvantaged compared to the central targets. Bifurcation space can be quite rich with interacting targets, and the exact geometry changes depending upon the radius of the target array from the central point.

\begin{figure}
    \centering
    \includegraphics[width=\linewidth]{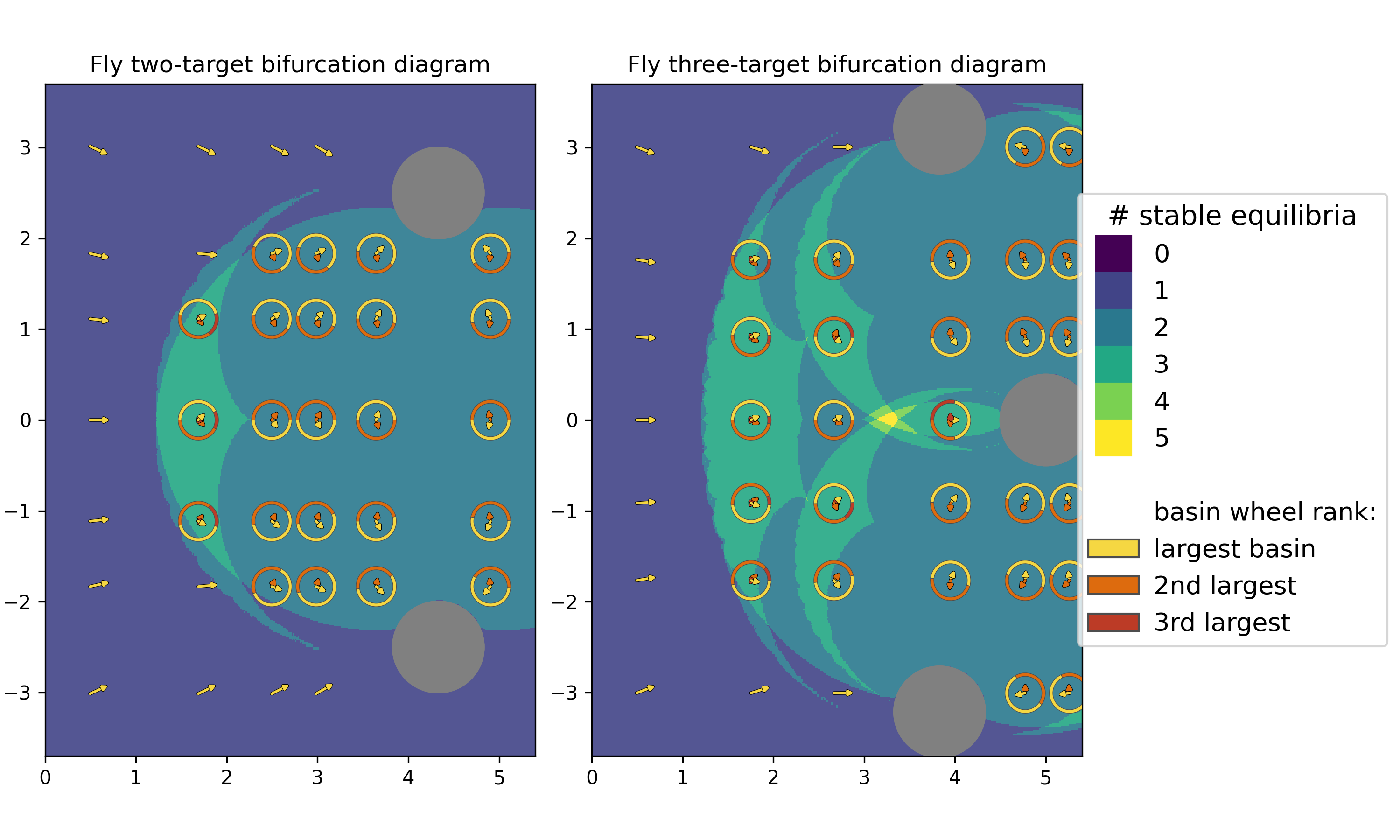}
    \caption{Bifurcation diagrams showing the number of self-consistent stable equilibria in $\theta$ as a function of spatial position for the two-target and three-target cases corresponding to the fly experiment in Sridhar \textit{et al.} \cite{Sridhar2021}. Directions of self-consistent equilibria are shown in a selection of locations, along with the approximate basins of attraction for an uncommitted observer ($R=1.5$) in cases of 2 or more stable equilibria. Parameterization was the same as in Fig. \ref{fig:walkers}.}
    \label{fig:fly_bifurcation}
\end{figure}

Fig. \ref{fig:fly_bifurcation} demonstrates our approach to quantifying the basin of attraction for an uncommitted observer at locations where there are more than one stable SC-equilibrium. Arrows in the middle of the annuli point in the SC-equilibrium, helping to identify if the equilibria correspond to target directions or consensus directions. Note that the arrows \textit{do not}, in general, point toward the center of their basin of attractino (see for example the annuli near the top right). Thus basins can be highly skewed in one direction or another. Near bifurcation points in $(x,y)$ space (boundaries of the colored regions denoting number of stable equilibria), the basin extent of one of the equilibria can be expected to shrink to zero. 

\clearpage

\bibliographystyle{unsrt} 
\bibliography{locust2026} 

\end{document}